\documentclass[aps,twocolumn,showpacs,floatfix]{revtex4}
\usepackage{graphicx}
\usepackage{dcolumn}
\usepackage{bm}
\usepackage[latin1]{inputenc}
\usepackage[colorlinks, citecolor=red, linkcolor=blue]{hyperref}
\usepackage{color}
\usepackage{multirow}

\makeatletter

\renewcommand*{\p@subsection}{}

\renewcommand*{\p@subsubsection}{}
\makeatother

\graphicspath{{./Figures/}}

\begin{document}

\preprint{}
\title{Experimental setup for the production of ultracold strongly correlated fermionic superfluids of $^{6}$Li}

\author{D. Hernández-Rajkov$^1$}
\author{J. E. Padilla-Castillo$^1$}
\author{M. Mendoza-López$^1$}
\author{R. Colín-Rodríguez$^1$}\altaffiliation[Present address: ]{Departamento de Física, Universidad Autónoma Metropolitana - Iztapalapa, Apartado Postal: 55-534, 09340, Ciudad de México, Mexico}
\author{A. Gutiérrez-Valdés$^1$}
\author{S. A. Morales-Ramírez$^1$}
\author{R. A. Gutiérrez-Arenas$^1$}
\author{C. A. Gardea-Flores$^1$}
\author{G. Roati$^2$}
\author{R. Jáuregui-Renaud$^1$}
\author{F. J. Poveda-Cuevas$^3$}
\author{J. A. Seman $^1$}\email{seman@fisica.unam.mx}

\affiliation{ }
\affiliation{$^1$ Instituto de Física, Universidad Nacional Autónoma de México, 01000 Ciudad de México, Mexico} 
\affiliation{$^2$ Istituto Nazionale di Ottica del Consiglio Nazionale delle Ricerche and European Laboratory for Nonlinear Spectroscopy (INO-CNR \& LENS), 50019 Sesto Fiorentino, Italy} 
\affiliation{$^3$ Cátedras CONACyT - Instituto de Física, Universidad Nacional Autónoma de México, 01000 Ciudad de México, Mexico}

\begin{abstract}

We present our experimental setup to produce ultracold strongly correlated fermionic superfluids made of a two-component spin-mixture of $^6$Li atoms. Employing standard cooling techniques, we achieve quantum degeneracy in a single-beam optical dipole trap. Our setup is capable of generating spin-balanced samples at temperatures as low as $T/T_F = 0.1$ containing up to $5 \times 10^4$ atomic pairs. We can access different superfluid regimes by tuning the interparticle interactions close to a broad magnetic Feshbach resonance. In particular, we are able to explore the crossover from the molecular Bose-Einstein condensate (BEC) to the Bardeen-Cooper-Schrieffer (BCS) superfluid regimes.

\end{abstract}


\maketitle

\section{Introduction}\label{sec:intro}

Quantum gases are macroscopic quantum many-body systems that represent a unique scenario to study quantum phenomena such as superfluidity and macroscopic quantum excitations \cite{Pethick}. Moreover, ultracold atoms have emerged as ideal quantum simulators of many-body phenomena, becoming effective testbeds of quantum Hamiltonians. Indeed, the combination of ultracold atoms and optical potentials has opened up a new way of studying condensed matter problems with unprecedented clarity \cite{Bloch}. This is possible thanks to the high level of controllability that quantum gases offer. The dimensionality and geometry of the system can be precisely modified by tailoring trapping potentials with laser light and magnetic fields. The thermodynamic properties of the gas, such as density, temperature, and volume can be easily manipulated. Full control of interparticle interactions is possible via magnetic Feshbach resonances \cite{Chin}. Even the quantum statistics of the system can be changed by choosing fermionic or bosonic atoms. These are very powerful tools that distinguish ultracold atomic gases from ordinary condensed matter systems. 

At ultralow temperatures, diluted gases composed of alkali-metal atoms interact predominantly through the $s$-wave scattering channel, since at such low energies, higher order collision channels are highly suppressed. In the case of ultracold bosons, Bose-Einstein condensation is possible and superfluidity emerges as long as the $s$-wave scattering length $a_s$ has a non-vanishing value \cite{Pethick}.

The case of ultracold identical fermions is strikingly different. In this case, $s$-wave scattering is also suppressed due to Pauli blocking, making these systems nearly non-interacting and, in consequence, they do not exhibit superfluid behavior even at zero temperature. The quantum degenerate state corresponds to an ideal Fermi gas, also known as Fermi sea \cite{Giorgini}. 

However, it is possible to introduce interactions into the system by creating a two-component spin mixture since Pauli blocking occurs only between identical fermions, while atoms in different spin states still interact via $s$-wave scattering. The absolute value of $a_s$ determines the interaction strength and its sign defines whether the interaction is effectively repulsive ($a_s > 0$) or attractive ($a_s < 0$). In fermionic systems, the interaction strength is usually described by introducing the dimensionless interaction parameter $(k_F\,a_s)^{-1}$, where $k_F$ is the Fermi wave vector \cite{Fermi}. Additionally, the existence of magnetic Feshbach resonances allows controlling the value of the scattering length practically at will by applying a constant magnetic field into the system. Therefore, varying the value of this external field makes the creation of different interaction regimes possible, from weakly to strongly interacting systems, from a repulsive to an attractive gas \cite{Chin}

\begin{figure} 
    \centering
    \includegraphics[width=\linewidth]{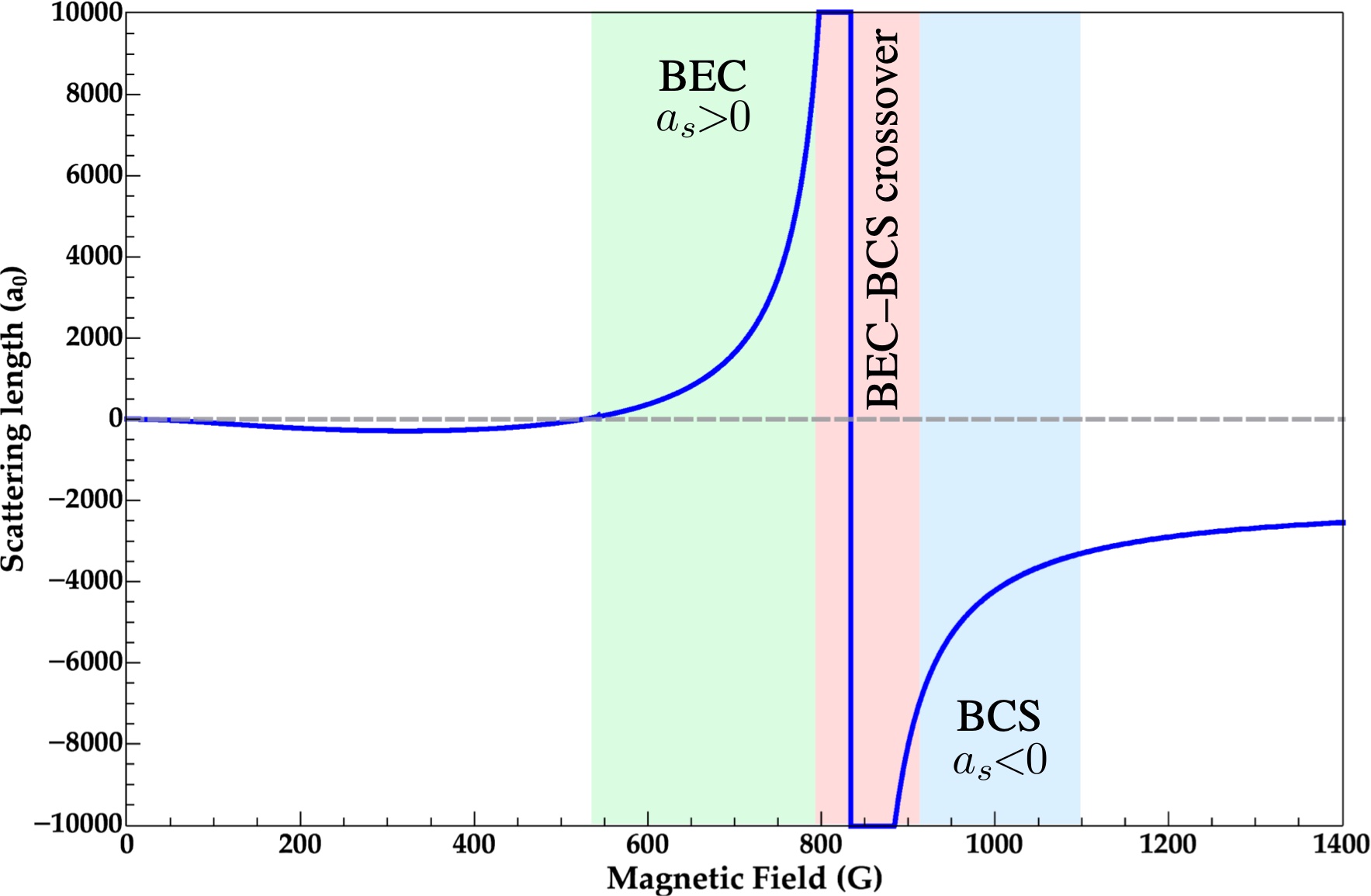}
    \caption{Feshbach resonance for the two lowest hyperfine Zeeman levels of $^6$Li. Different superfluid regimes are possible depending on the value of the scattering length $a_s$.}\label{fig:feshbach}
\end{figure}

A very important consequence of having such control on interatomic interactions is the possibility of creating different types of bound states among the atoms. For repulsive interactions, $(k_F\,a_s)^{-1} > 0$, a molecular bound state exists in the interaction potential. In this case, it is possible to associate diatomic molecules composed by two identical atoms which, consequently, will exhibit bosonic statistics making possible the emergence of Bose-Einstein condensation of tightly bound molecules \cite{Jochim,Zwierlein,Greiner}. On the other hand, for attractive interactions, $(k_F\,a_s)^{-1} < 0$, the corresponding bound state occurs in momentum space due to many-body correlations at the Fermi surface, giving rise to loosely bound Cooper-like pairs whose behavior is well described by the BCS theory \cite{Cooper, BCS1, BCS2}. In this way, the Feshbach resonance allows to continuously transit from the BEC to the BCS regimes through the so-called BEC-BCS crossover \cite{Bourdel,Chin2004,Zwerger}. The intermediate regime in which the scattering length diverges, $(k_F\,a_s)^{-1} = 0$, known as unitary limit, is particularly intriguing  because the system is strongly interacting and strongly correlated giving rise to interesting physics. Indeed, being a universal regime, physics across the BEC-BCS crossover is interesting due to its relationship with other important phenomena such as high-$T_c$ superconductivity \cite{Chen} and other strongly correlated superfluids such as neutron stars or quark-gluon plasma \cite{Adams, Salasnich,Wyk}. Figure~\ref{fig:feshbach} shows the Feshbach resonance for the case of the two lowest hyperfine Zeeman levels of $^6$Li, corresponding to the two Zeeman components of the absolute ground state $2^2S_{1/2}$ $F=1/2$, $m_F = -1/2$ and $m_F = +1/2$, which we denote as $| 1 \rangle$ and  $| 2 \rangle$ respectively. Figure~\ref{fig:feshbach} also specifies the different superfluid states for the different interaction regimes. This Feshbach resonance is particularly broad, having a width of  the order of 300\,G, enabling a very fine and precise control of the scattering length. 

In this paper we describe our newly built setup to produce ultracold atomic gases composed by fermionic $^6$Li atoms in a balanced spin-mixture of the states $| 1 \rangle$ and  $| 2 \rangle$. We employ a standard Zeeman slower to decelerate an atomic beam coming from an effusive oven. The decelerated atoms are trapped and further cooled down in a magneto-optical trap in which subsequently sub-Doppler cooling is used \cite{Metcalf}. These laser cooled atoms are transferred into a single-beam optical dipole trap \cite{Metcalf,Grimm}. Finally, quantum degeneracy is achieved by runaway evaporation. We produce ultracold samples in the different superfluid regimes across the BEC-BCS crossover containing about $5 \times 10^4$ pairs at an approximate temperature of $T/T_F = 0.1$. 

The article is divided as follows. In Section~\ref{sec:setup} we provide details of our experimental setup, this includes the ultra-high vacuum system, the laser system, the magnetic field generation system, and the creation of conservative potentials. Section~\ref{sec:results} is devoted to the procedures used to cool down the gas to quantum degeneracy: laser cooling and evaporative cooling techniques, as well as details on the production of a superfluid sample in different interacting regimes. Finally, in Section~\ref{sec:conclusions}, we present our conclusions and future perspectives.

\section{Experimental Setup}\label{sec:setup}

	\subsection{The ultra-high vacuum system}\label{sec:UHV}

\begin{figure*}[t]
    \centering
    \includegraphics[width=0.9\linewidth]{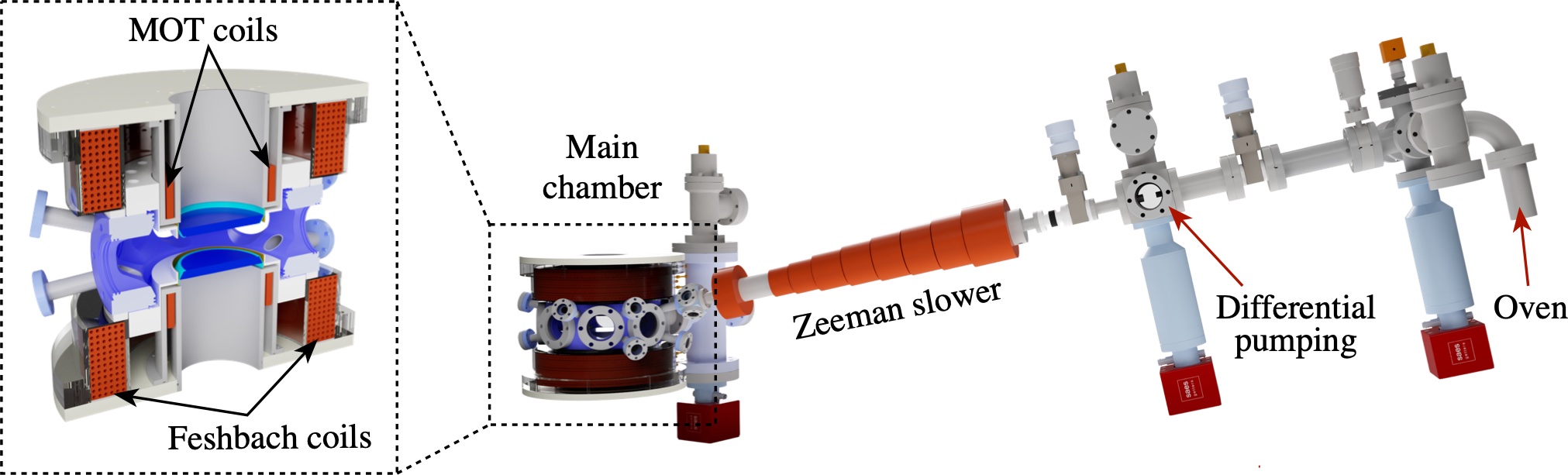}
    \caption{Scheme of the ultra-high vacuum system including the Zeeman and the Feshbach coil systems. On the left we show a cut of the main chamber, exhibiting the distribution of the Feshbach and MOT coils. See text for details.}
    \label{fig:UHV}
\end{figure*}

Our ultra-high vaccum (UHV) system is divided in three main sections, namely (\textit{i}) the effusive oven; (\textit{ii}) the differential pumping stage, and (\textit{iii}) the Zeeman slower system and the main chamber where the sample is produced and the experiments performed. Each of these sections is pumped by a 200~l/s pumping system composed by a combination of an ion pump and a non-evaporable getter (model NEXTorr\textsuperscript{\textregistered} D 200-5 from SAES getters Inc.). Figure~\ref{fig:UHV} shows a scheme of our UHV, including a detailed cut of our main chamber.

The effusive oven consists of a cylindrical recipient which contains 5\,gr of purified $^6$Li. The oven is heated to a temperature of 450$^\circ$C, at this temperature the vapor pressure of lithium is about $1\times 10^{-4}$~Torr \cite{Nesmeyanov}. The oven is connected to the rest of the UHV system through a 4\,mm diameter nozzle. The vapor produced in the oven passes through this nozzle generating an atomic beam that propagates through the rest of the system. We estimate that the atomic flux of the beam effusing out from the nozzle is of the order of $6 \times 10^{15}$\,atoms/s \cite{flux}.

The pressure right after the nozzle reaches a value well above $10^{-9}$~Torr, which is too high for producing quantum degenerate samples. To keep a sustained pressure difference between the oven and the region in which experiments are performed, we have placed a differential pumping stage which consists of two aligned tubes separated by 25\,mm from each other:  the first one, facing the oven, has a 4.6~mm inner diameter and a second one, facing the Zeeman slower, is 7.7~mm inner diameter. This scheme is designed to keep a pressure difference across the differential pumping stage up to five orders of magnitude. In this way, the pressure in the main chamber is of the order of $10^{-11}$~Torr.

The main chamber is connected to the differential pumping stage by a 56\,cm long and 16.5~mm inner diameter tube. Around this tube there is a conical solenoid which is used to create a spatially inhomogeneous magnetic field which is required to implement a Zeeman slower system (more details are provided in Section~\ref{sec:mag_Zeeman}).

Finally, the main chamber is a stainless steel custom-made octagon chamber from Kimball Physics Inc. This chamber contains eight CF40 viewports on its sides; two CF100 vertical viewports, and ten CF16 viewports connected to the chamber by arms extruded from it at an angle of 13$^\circ$ from the horizontal plane. The Zeeman slower tube is connected to the main chamber by one of these arms while the Zeeman slower laser beam enters through another one diametrically opposed with respect to the center of the chamber. All viewports have anti-reflection coating for all the wavelengths used in our experiment (532\,nm, 671\,nm and 1064\,nm).

We have placed on both CF100 flanges reentrant viewports of high optical quality whose inner face is very close to the atoms, at a distance of only half an inch. This opens the possibility of building a large numerical aperture optical system to produce high resolution images of the sample.

	\subsection{Laser system}\label{sec:optics}
		
		\subsubsection{Optical cooling scheme}\label{sec:cool_optics}

To implement the different laser cooling techniques in our experiment we use both the $D_2$ and $D_1$ optical transitions of $^6$Li \cite{Das}. We use two separate diode lasers, one for each line. The emission frequency of these lasers is locked-in into an atomic reference using a spectroscopy cell containing 5\,gr of purified $^6$Li heated at 320ºC where we implement standard saturated absorption spectroscopy (SAS) \cite{Letokhov}. We use the same cell to lock-in both lasers.

\begin{figure}[ht]
    \centering
    \includegraphics[width=0.95\linewidth]{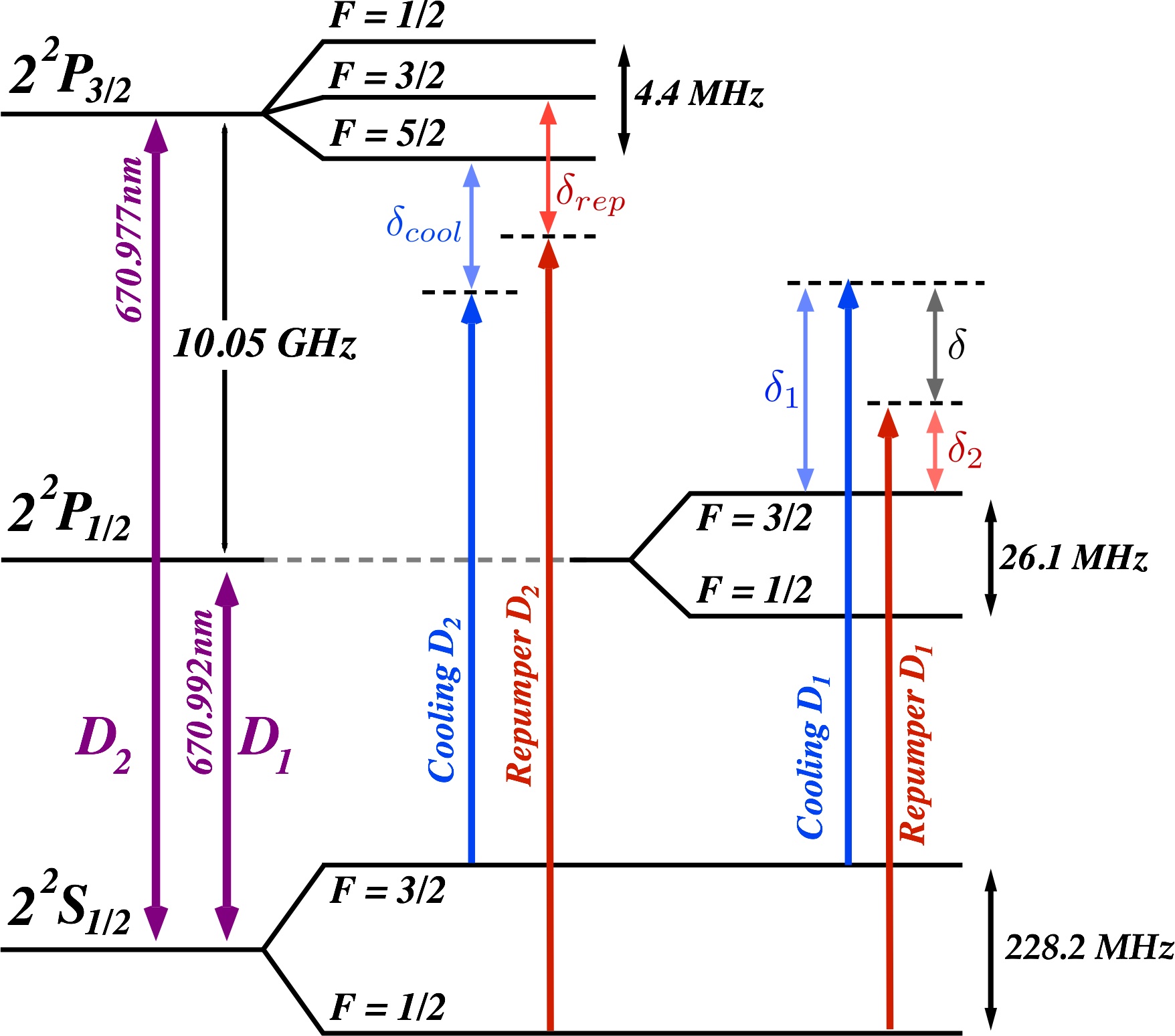}
    \caption{Level scheme (not to scale) for $^6$Li showing (left) the $D_2$ and (right) the $D_1$ hyperfine structures and the transitions used for the laser cooling processes. See text for details.}
    \label{fig:D1D2}
\end{figure}

In the experiment, the $D_2$ line is used first to implement the magneto-optical trap (MOT) and later, for the optical molasses cooling stage, while the $D_1$ line is subsequently used to apply a sub-Doppler cooling stage. The natural linewidth of both lines is $\Gamma = 2\pi \times 5.87$\,MHz \cite{McAlexander}. The main optical frequencies employed in our experiment are shown in Figure~\ref{fig:D1D2}. 

\begin{figure*}[ht]
    \centering
    \includegraphics[width=0.9\linewidth]{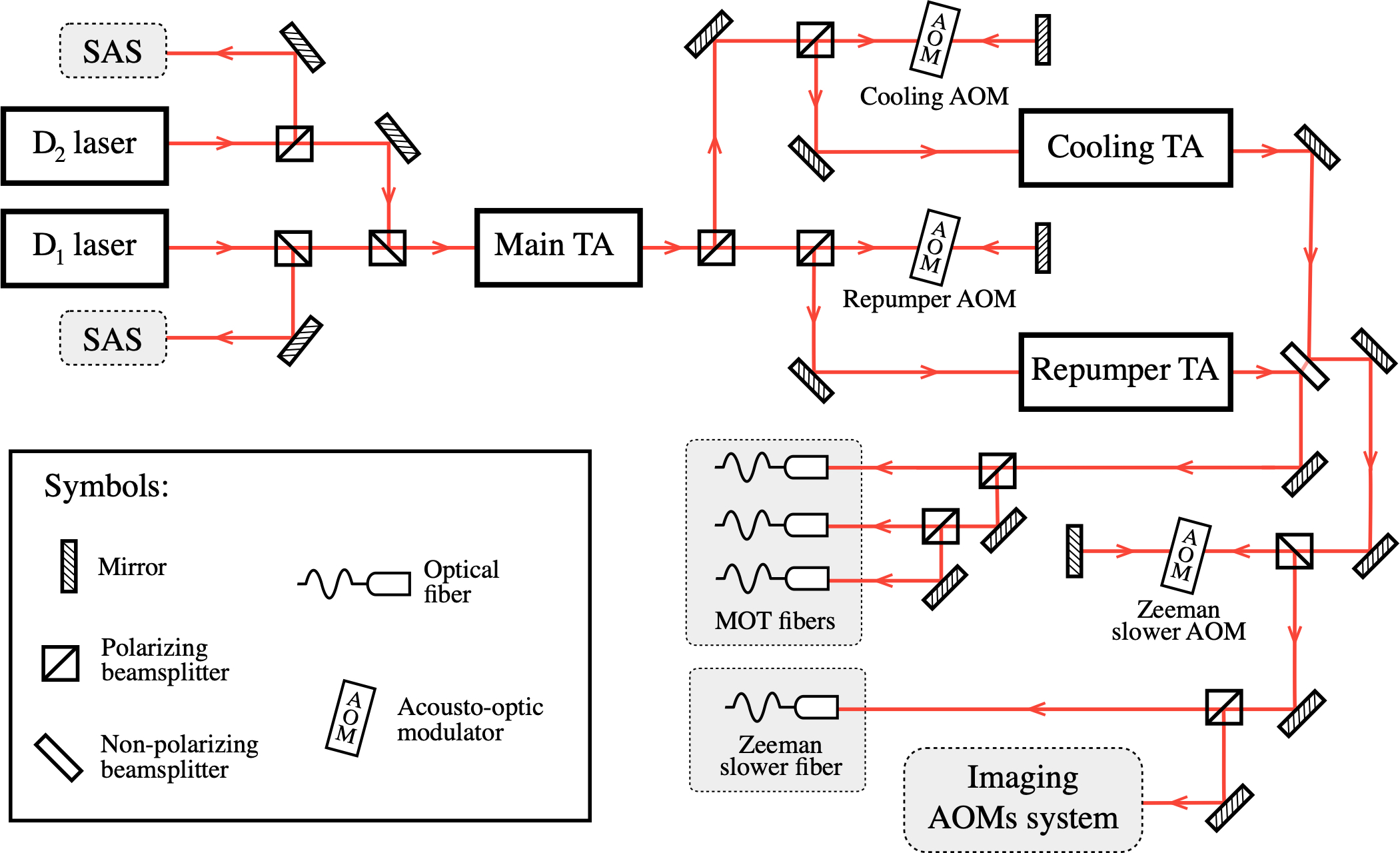}
    \caption{Simplified scheme of the laser cooling and imaging optical setup showing the main features of the system. Lenses and waveplates have been omitted for clarity. See text for details.}
    \label{fig:optical}
\end{figure*}

To produce the $D_2$ frequencies we have an extended-cavity diode laser (cat-eye configuration, model CEL002 from MOGLabs) which pumps an optical tapered amplifier (model MOA002 also from MOGLabs). We divide the amplified beam into two beams and independently shift the frequency of each of them using two different acousto-optic modulators (AOM). Next, each one of these beams pumps another tapered amplifier and in this way we generate two high power beams (power of $\sim$500\,mW each) at a wavelength of approximately 670.9~nm with a frequency difference between them of 228.2\,MHz, which corresponds to the hyperfine splitting of the ground state $2^2S_{1/2}$ of $^6$Li. One of these beams, the one with lower frequency, corresponds to the cooling frequency which is red-detuned from the $2^2S_{1/2} (F=3/2) \rightarrow 2^2P_{3/2}$ transition by $8.5\,\Gamma$ (about 50\,MHz). The second beam is used as repumper frequency and is red-detuned from the $2^2S_{1/2} (F=1/2) \rightarrow 2^2P_{3/2}$ transition by $8.5\,\Gamma$. Note that we do not specify the hyperfine level of the excited state $2^2P_{3/2}$ in either the cooling nor repumper frequencies. This is because the energy levels of these hyperfine states are too close together, their separation is less than $\Gamma$, and therefore we can not resolve them in our spectroscopy cell. We superimpose both beams using a 50:50 non-polarizing beam splitter which, hence, produces two beams with the same power, each one carrying both cooling and repumper frequencies. One of these beams is used to generate the light for the MOT. To do so, we subsequently divide it into three equally powered beams and couple each one into a polarization maintaining optical fiber which brings the light directly to the experimental region. The second beam coming from this 50:50 beam splitter is additionally red-shifted by $76\,\Gamma$ using an additional AOM. In this way we produce the Zeeman slower beam (which also arrives into the experiment by a polarization maintaining optical fiber). This large frequency shift is chosen to match the Doppler and Zeeman shifted $D_2$ line levels of the fast atoms coming out from the oven at the beginning of the Zeeman slower coil (where its magnetic field is maximum).

On the other hand, to implement the $D_1$ sub-Doppler cooling stage, we require the two frequencies that are shown in the right side of Figure~\ref{fig:D1D2}. We have a cooling frequency blue-detuned from the transition $2^2S_{1/2} (F=3/2) \rightarrow 2^2P_{1/2} (F'=3/2)$ by $5\,\Gamma$ (about 30\,MHz), and a repumper frequency blue-detuned from the transition $2^2S_{1/2} (F=1/2) \rightarrow 2^2P_{1/2} (F'=3/2)$ also by $5\,\Gamma$. Note that we never use both $D_2$ and $D_1$ lines at the same time and that the cooling and repumper frequencies of the $D_1$ line are, evidently, also separated by 228.2\,MHz. This enables us to use exactly the same optical setup that we employed to generate the $D_2$ frequencies. We have a second diode laser (same model than before) whose light we superimpose, using a polarizing beam splitter cube, onto the very same optical path of the $D_2$ line laser.  Finally, the $D_1$ light reaches the sample using the same optical fibers that were used for the MOT. A simplified sketch of our laser setup is presented in Figure~\ref{fig:optical}.

As can be seen, we essentially set all the required frequencies using three AOMs in double-pass configuration \cite{Donley}. These AOMs are also used to dynamically change the frequency of these beams and implement the $D_2$ optical molasses and $D_1$ sub-Doppler cooling stages, as explained in Section~\ref{sec:laser_cool}.

		\subsubsection{Generation of probing light}\label{sec:diag_optics}

The most important diagnostic tool in cold atoms experiments is imaging the samples using laser light. In our case the preferred technique is absorption imaging due to its simplicity and reliability \cite{KetterleBose,KetterleFermi}.

Absorption imaging consists in probing the sample using a collimated laser beam whose frequency is resonant to some atomic transition. To produce the image, we pulse this light on the atoms during a short time (of the order of $5~\mu$s). The atoms will absorb some of the light, generating an absorption profile on the beam. Finally, the light is collected by a telescope that creates an image of such absorption profile on a CCD camera (model MANTA G-145 NIR from Allied Vision Technology GmbH). The density profile of the gas can be extracted from this image \cite{KetterleBose}.

In our experiment we want to produce samples at different interaction regimes across the BEC-BCS crossover. This is done by applying an external magnetic field that changes the value of the scattering length by means of a Feshbach resonance. This magnetic field, in turn, will also cause a Zeeman splitting on the electronic levels of the atoms. Hence, probing the atoms at different interaction regimes poses the necessity of generating different light frequencies to keep the imaging light resonant with the atoms. 

To do so, we use the Zeeman slower beam which already has a considerable shift of $-76\,\Gamma$. We divert a fraction of this beam using a polarizing beam splitter before it is coupled into the Zeeman slower optical fiber, as shown in Figure~\ref{fig:optical}. Next, this diverted beam passes through additional AOMs that will further shift the frequency to match it to the specific magnetic field in which we want to probe the atoms. This configuration of AOMs allows to tune the frequency of the probing light at different values within the range from $0$ to $-220 \Gamma$ from the $D_2$ transitions. In this way, we are able to produce images at practically any magnetic field from 200~G to 1200~G and also at the vicinity of zero magnetic field. In this way, as can be seen in Figure~\ref{fig:feshbach}, we can image the sample in all the superfluid regimes across the BEC-BCS crossover, and also at weakly interacting regimes in which the gas is simply a Fermi degenerate gas but not a superfluid.

Finally, it is important to mention that the magnetic field used to access the BEC-BCS crossover is high enough to ensure that the hyperfine splitting of the atoms is well within the Paschen-Back regime, where the separation between the $|1\rangle$ and $|2\rangle$ states remains almost constant at approximately 76\,MHz. For this reason, we can probe both spin states in any magnetic field through the Feshbach resonance.

	\subsection{Magnetic field generation system}\label{sec:magnetic}

We employ three different sets of coils to generate all the required magnetic fields to trap and manipulate the atoms. We describe each of them in the following sections.

		\subsubsection{Zeeman slower magnetic field}\label{sec:mag_Zeeman}

We use a Zeeman slower stage to decelerate the atomic beam coming out from the effusive oven. As mentioned in the previous section, the detuning of the Zeeman slower laser beam is $\delta_{Z} = -76\Gamma$ for both cooling and repumper frequencies. The corresponding magnetic field along the direction of propagation of the atomic beam (direction $z$) is designed to decelerate atoms with velocities up to $v_0 = 960$\,m/s at a constant deceleration $a \approx \hbar k \Gamma / 2m$ \cite{a} through the formula \cite{Metcalf}:

\begin{eqnarray}\label{eq:BZ}
B(z) = \frac{\hbar}{\mu_B}\left( \delta_Z + k\sqrt{v^2_0 - 2az} \right),
\end{eqnarray}
where $\mu_B$ is the Bohr magneton and $k$ is the wavevector of the cooling frequency of the slower light. In this formula we only consider the cooling frequency since it is the one responsible for the deceleration of the atoms.

\begin{figure}
    \centering
    \includegraphics[width=0.95\linewidth]{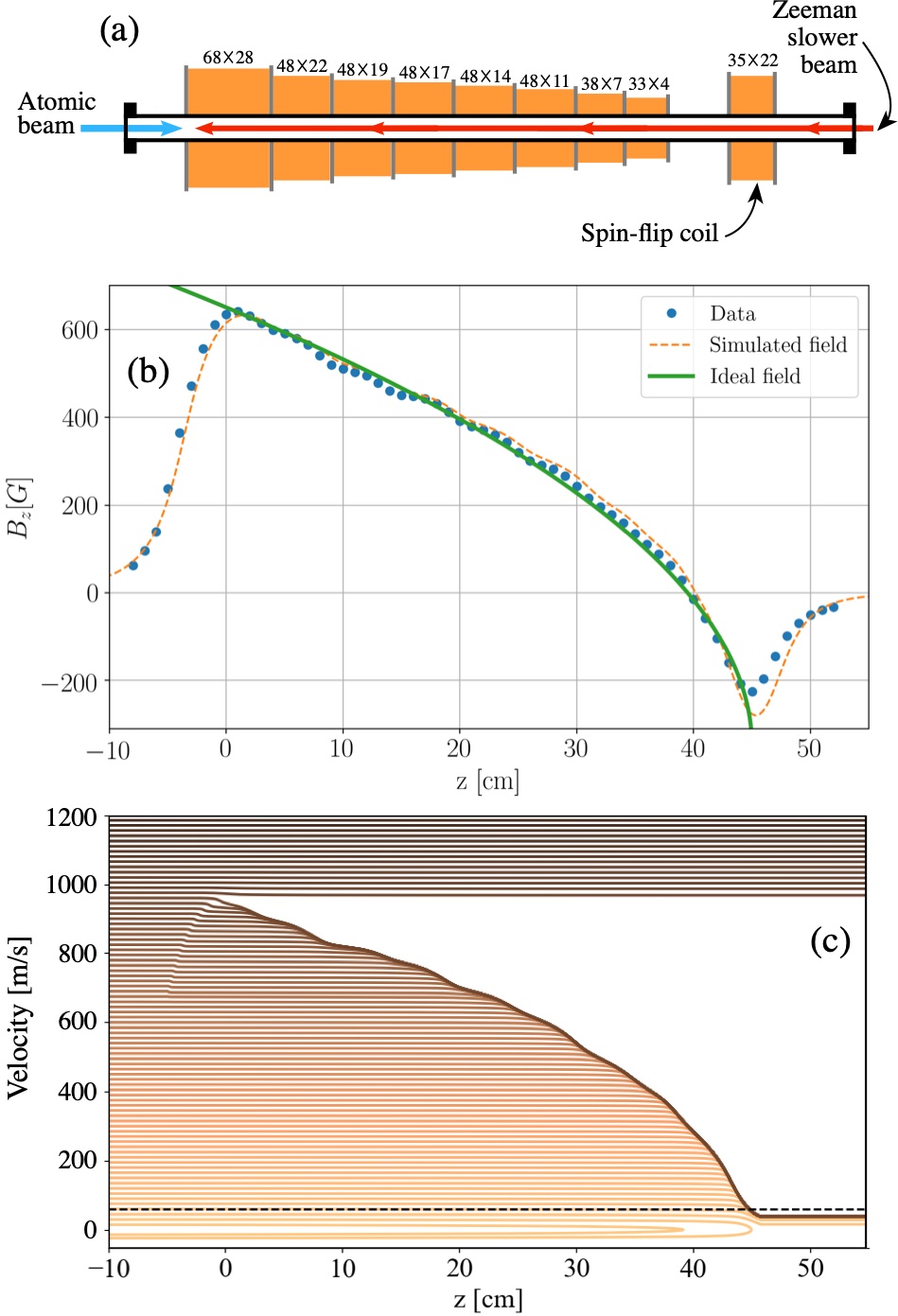}
    \caption{(a) Scheme of the coils of our Zeeman slower, the number of windings of each coil is indicated in the format $H \times V$, where $V$ denotes the number of layers in the vertical direction and $H$ provides the number of turns in each layer. (b) Axial component of the magnetic field generated along the Zeeman slower, the blue dots are the experimental data, the orange dashed line is the simulated field for this coil configuration and the solid green curve is the ideal magnetic field obtained through Eq.~(\ref{eq:BZ}). The data uncertainty is of 1\% however the corresponding error bars are not visible at this scale. (c) Evolution of the speed of the atoms propagating through the Zeeman slower, the dashed horizontal line indicates the capture velocity of the MOT. }
    \label{fig:slower}
\end{figure}

The mean velocity of the atoms coming out from the oven is $\bar{v} \simeq 1540$\,m/s, which is higher than the maximum velocity $v_0$ we can decelerate in our slower. This means that, in the best case, we can only slow down about 20\% of the atoms. This is not a problem since the flux of atoms effusing from the oven is very large, about $6\times10^{15}$\,atoms/s, so we can still efficiently load our magneto-optical trap.

The magnetic field of the slower is generated by a succession of eight size-decreasing coils connected in series and an extra ninth coil at the end of the slower in which the current circulates in opposite direction, inverting in this way the direction of the magnetic field. This is known as ``spin-flip configuration'' \cite{Natarajan,Guevara}. All nine coils are wound directly onto the slower UHV tube using 1\,mm diameter cooper wire. The coils are held together using a thermal conducting and electric insulating ceramic epoxy (Duralco\textsuperscript{TM} 128). The total current passing through each coil is of the order of 2.0\,A to generate a field which goes from a maximum around $600$\,G to a minimum of about $-250$\,G. 

Figure~\ref{fig:slower}\,(a) shows a scheme of the coil configuration of our Zeeman slower. Figure~\ref{fig:slower}\,(b) presents the generated magnetic field. Finally, Figure~\ref{fig:slower}\,(c) exhibits the calculated velocity profile of the decelerated atoms through their propagation along the slower.

		\subsubsection{Magnetic quadrupole}\label{sec:mag_MOT}

To produce the magneto-optical trap we use a quadrupole magnetic field whose axial gradient at the center of the trap is $\partial	 B_z(z)/\partial z |_{z=0} \simeq 28$~G/cm. This field is  generated by two small coils of $6 \times 4$ windings connected in anti-Helmholtz configuration. Each of these coils is mounted in a cylindrical water-cooled support to prevent them from heating. This support is made of Teflon\textsuperscript{TM}, which is a non-magnetic and insulating material, this prevents eddy currents from being induced in it when the quadrupole field is abruptly varied. Both supports are mounted inside the reentrant viewports of the main chamber, along the vertical direction. The coils are wound with strip-shaped copper wire of 4\,mm $\times$ 1\,mm and held together with ceramic epoxy (Duralco\textsuperscript{TM} 128). Figure~\ref{fig:UHV}\ shows the position of these coils in the main chamber.

		\subsubsection{Feshbach resonance magnetic field}\label{sec:mag_Fesh}

As already mentioned, one of the important advantages of ultracold lithium gases is the possibility of controlling interatomic interactions with a high degree of precision by means of a Feshbach resonance \cite{Chin}. $^6$Li presents several Feshbach resonances whose characteristics depend on the internal state of the interacting atomic pair. We will use the resonance between states $| 1 \rangle$ and $| 2 \rangle$, shown in Figure~\ref{fig:feshbach}, centered at 832\,G. So we need an extra set of coils that are able to produce an uniform magnetic field with any value from zero to 1000\,G in order to have full control of all interaction regimes. To do so we have a pair of coils connected nearly at Helmholtz configuration. We deliberately move slightly away from the Helmholtz configuration so the magnetic field presents a small curvature, which will be useful to confine the atoms along the weak direction of our optical dipole trap; right at the Feshbach resonance, at 832\,G, this curvature along the coils axis direction is $B''_z(0) = 6.2$\,G/cm$^2$, while the corresponding magnetic gradient is nearly zero (see Section~\ref{sec:ODT} \, for more details). 

The Feshbach coils are made by 4\,mm square section copper wire. This wire is hollow, with an internal diameter of 2\,mm, which allows cooling the coil by circulating cold water inside the wire. These coils were fabricated by the company Oswald Elektromotoren GmbH and each of them is embedded in an insulating resin that avoids the induction of undesired eddy currents. We can circulate a current above 200~A without noticing any significant heating of the coils. This thermal stability together with a PID feedback loop makes possible to produce magnetic fields with a stability of one part in $10^4$. We place these coils along the vertical direction, colinear to the quadrupole field coils. Figure~\ref{fig:UHV} shows each of the employed set of coils and their position in the experimental setup.

	\subsection{Conservative trapping potential}\label{sec:ODT}
	
We produce the quantum degenerate fermionic system in a conservative trap generated by the combination of an optical potential and a magnetic curvature.
	
The optical potential consists in a far red-detuned single-beam optical dipole trap (ODT) created by focusing a gaussian infrared laser beam \cite{Grimm}. We use a single mode ytterbium-doped fiber laser from IPG Photonics Corp. (model YLR-200-LP) which delivers up to 200\,W of continuum linearly polarized infrared light at $\lambda = 1070$\,nm. The beam of this laser is coupled into an acousto-optic modulator and the first diffracted order is used to produce the optical trap. We use a quartz crystal AOM that withstands very high intensities, above 1\,GW/cm$^2$, from the company Gooch~\&~Housego (model \textit{I-M080-2C10G-4-AM3}). The diffraction efficiency of this AOM and, consequently, the power of the ODT, is manipulated by controlling the amplitude of the RF signal that drives the modulator using an external analog signal. To stabilize the power of this diffracted order, we employ a PID circuit driven by the signal of a photodiode (Thorlabs, model DET36A) that detects the small fraction of the light of this beam that is transmitted by a 99.9\% reflection mirror.

Next, we collimate the beam at a diameter of approximately $D = 5.5$\,mm and finally use a $f = 40$\,cm focal length lens to focus the light on the atoms. The beam waist at focus is $\mbox{w}_0 = 2 \lambda f / \pi D \simeq 50~\mu$m, which corresponds to a Rayleigh length of $z_R = \pi\mbox{w}^2_0/\lambda = 7.34$\,mm.
	
The trap frequencies of this single beam ODT along the radial and axial directions are given, respectively, by \cite{Grimm}

\begin{eqnarray}\label{eq:ODT_freqs}
\omega_{r_{ODT}} = \sqrt{\frac{4U_0}{m\mbox{w}^2_0}} \ \ \ \mbox{and} \ \ \ \omega_{z_{ODT}} = \sqrt{\frac{2U_0}{m\mbox{z}^2_R}},
\end{eqnarray}
where $U_0$ is the depth of the trapping potential and it is given by \cite{Grimm}

\begin{eqnarray}\label{eq:U0}
U_0 = \left( \frac{3\pi c^2 \Gamma}{2\omega^3_0\Delta\omega} \right) I_0,
\end{eqnarray}
where $\omega_0$ is the frequency of the lowest energy optical dipole transition of the $^6$Li atom, which corresponds to the $D_1$ transition, $\Gamma$ is the natural linewidth of such transition and $\Delta\omega = \omega_0 - \omega$ is the detuning between such transition and the ODT laser frequency $\omega$. Finally, $I_0$ is the intensity of the ODT beam at focus, $I_0 = 2P/\pi\mbox{w}^2_0$, where $P$ is the power of the ODT laser.

This trap provides a tight confinement along the radial direction of the beam, however, along the axial (or propagation) direction it is very weak. For instance, at the end of the evaporative cooling where the power of the ODT laser is approximately $P = 35$\,mW (see Section~\ref{sec:evap_cool} ), the radial and axial frequencies of the optical trap respectively are $\omega_r = 2\pi \times 163$~Hz and $\omega_z = 2\pi \times 0.94$~Hz, which would provide an extremely elongated sample with an aspect ratio larger than 1:160.
	
For this reason, we add to the optical potential a magnetic curvature that provides a better confinement along the axial direction. As mentioned in Section~\ref{sec:mag_Fesh}\, , we produce such curvature by setting the Feshbach coils slightly off the Helmholtz configuration. In this way, we create a saddle-point magnetic potential of the form \cite{KetterleFermi,Bergeman}

\begin{eqnarray}\label{eq:mag_field}
U_{mag}(r,z) = \frac{m}{2}\left(\omega^2_{z_{mag}}z^2 - \omega^2_{r_{mag}}r^2 \right),
\end{eqnarray}
where the trap frequencies are determined by the curvature of the field component along the corresponding direction, i.e. $\omega^2_{z_{mag}} = \mu B''_z(0)/m$ and $\omega^2_{r_{mag}} = \mu B''_r(0)/m$, being $m$ the mass of $^6$Li atom and $\mu$ the magnetic moment of the trapped state which, in general for the ground state of alkali atoms, is of the order of the Bohr magneton, $\mu \sim \mu_B$.

Note from Eq.~(\ref{eq:mag_field}) that along the radial direction we have an ``anti-curvature'' which will tend to deconfine the atoms along that direction. The total frequencies of our hybrid trap will be given simply by 

\begin{eqnarray}\label{eq:hybrid_freqs}
\omega_{r} = \sqrt{\omega^2_{r_{ODT}} - \omega^2_{r_{mag}}} \  \mbox{and}  \  \omega_{z} = \sqrt{\omega^2_{z_{ODT}} + \omega^2_{z_{mag}}}.
\end{eqnarray}

In our experiment, once the quantum sample is produced, we have that the radial optical confinement is much larger than the magnetic one ($\omega_{r} \approx \omega_{r_{ODT}}$), and, vice versa, along the axial direction the confinement is dominated by the magnetic component ($\omega_z \approx \omega_{z_{mag}}$).

The axial curvature generated for the fields used to access the BEC-BCS crossover is of the order of $B''_z(0) = 6.2$\,G/cm$^2$ which, superimposed to the ODT potential, translates into a total axial frequency of $\omega_z \simeq 2\pi \times 11$\,Hz. In this way we obtain a cigar-shaped quantum sample whose aspect ratio, of the order of 1:15, is appropriate for our goals.

\section{Methods and Results}\label{sec:results}

In the following sections we provide details on the experimental procedures employed to produce the ultracold samples. 

In a very general way, the production of the quantum sample can be divided into two main processes: \textit{(i)} an initial laser cooling stage mediated by absorption and reemission of light, explained in Section~\ref{sec:laser_cool}\, , and \textit{(ii)} transference into a conservative potential and cooling by forced evaporation, presented in Section~\ref{sec:evaporation}\,

	\subsection{Implementation of laser cooling technique}\label{sec:laser_cool}

In this first cooling process we are able to produce atomic samples at temperatures as low as 40~$\mu$K containing $4.5 \times 10^8$ atoms with a density of the order of $4.5 \times 10^{9}$ atoms/cm$^3$, which correspond to a phase-space density of about $6.6~\times~10^{-6}$. We provide details on the laser cooling procedure in the following sections.

		\subsubsection{Zeeman slower and magneto-optical trapping}\label{sec:MOT}
	
\paragraph*{\textbf{Zeeman slower operation:}}

 The quantum sample generation process starts by heating the lithium sample contained in the oven of our UHV system to 450$^{\circ}$C. This generates a high temperature atomic beam that propagates through the UHV system towards the main chamber. The atoms of this beam undergo a first cooling process as they are decelerated by our Zeeman slower. Along the slower, a laser beam propagates in the opposite direction to the atomic beam. This laser carries two different frequencies, both of them red-detuned by 76\,$\Gamma$ ($\sim$446~MHz) from the cooling and repumper transitions of the $D_2$ line. Each of these frequencies has a power of 40\,mW and carries positive circular polarization $\sigma^+$. In this way, we are able to decelerate all the atoms from velocities classes under 960\,m/s to speeds of the order of 40\,m/s, well below the 60\,m/s capture velocity of the MOT, as shown in Figure~\ref{fig:slower}(c).

We found that controlling independently the electric current of the spin-flip coil provides better results. Best results are obtained using a current of 2.0\,A for the spin-flip coil and 2.9\,A for all other coils, which optimize the number of loaded atoms into the MOT and minimize the corresponding loading time.

\paragraph*{\textbf{Loading of the magneto-optical trap:} }

\begin{figure}
    \centering
    \includegraphics[width=0.95\linewidth]{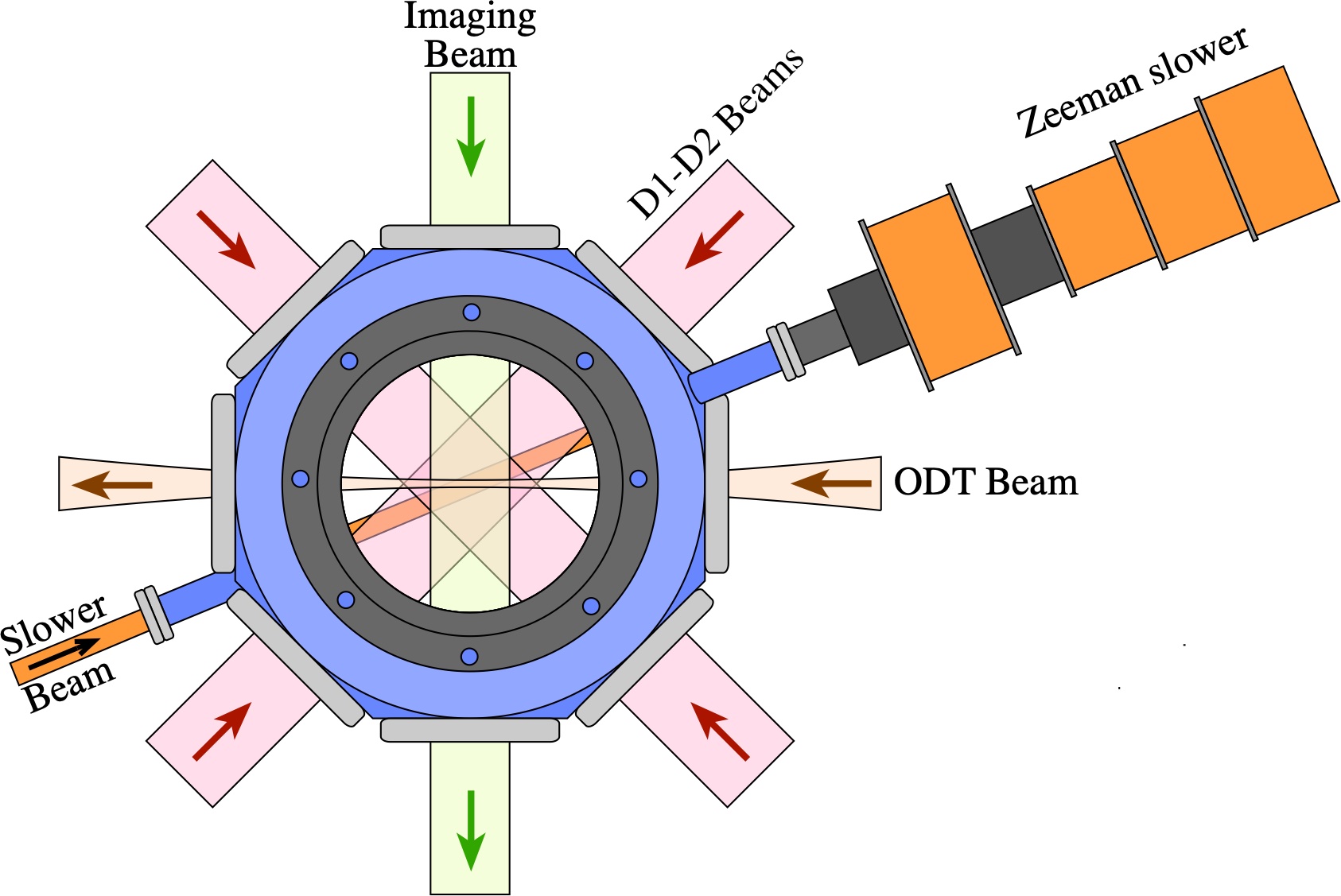}
    \caption{Top view scheme of the main chamber, showing the configuration of the MOT beams (D$_1$ and D$_2$ beams), the imaging beam, the Zeeman slower beam and the ODT beam. MOT and Feshbach coils were omitted for clarity. The third pair of MOT beams is perpendicular to the plane of this scheme and, hence, not shown.}
    \label{fig:Beams}
\end{figure}

The decelerated atoms arrive into the main chamber where we capture them and further cool them in a magneto-optical trap (MOT) \cite{Metcalf}. To implement the MOT we use three retroreflected mutually perpendicular laser beams with a diameter of $D = 2.3$\,cm, as shown in Figure~\ref{fig:Beams}. 

\begin{figure}
    \centering
    \includegraphics[width=\linewidth]{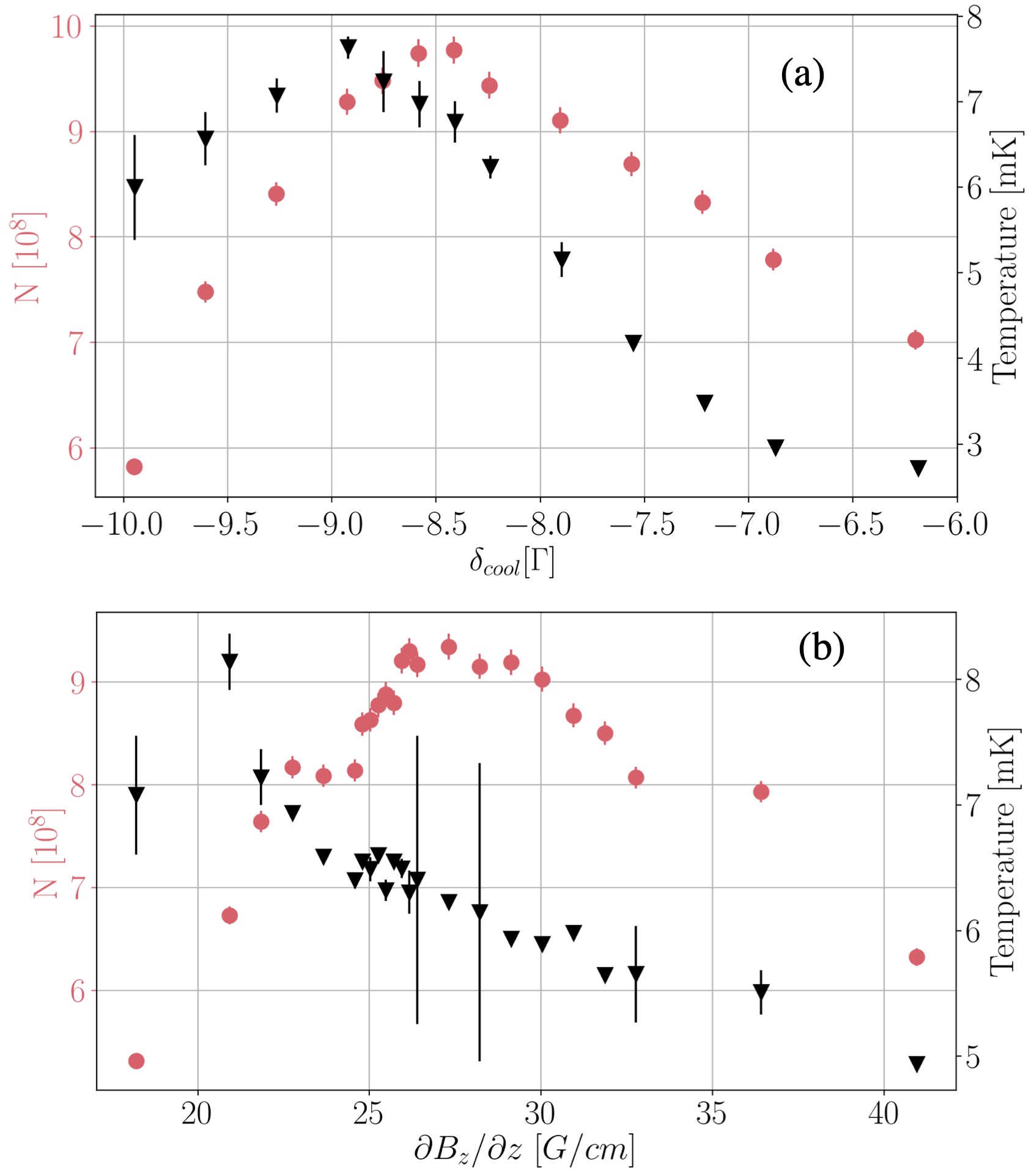}
    \caption{Number of atoms $N$ (red dots) and temperature $T$ (black triangles) of the atoms of the MOT as a function of (a) the detuning of the cooling light and (b) the axial gradient of the quadrupole magnetic field. In these plots, the error bars correspond to one standard deviation of ten independent measurements. }
    \label{fig:MOT}
\end{figure}

The MOT beams carry two frequencies: a cooling frequency, red-detuned from the $2^2S_{1/2} (F=3/2) \rightarrow 2^2P_{3/2}$ transition, and a repumper frequency, red-detuned from the $2^2S_{1/2} (F=1/2) \rightarrow 2^2P_{3/2}$ transition. We use the standard $\sigma_{+}/\sigma_{-}$ polarization configuration. We determine the value of the detunings by maximizing the number of atoms $N$ loaded into the MOT and by trying to keep the temperature of the sample $T$ as low as possible. Figure~\ref{fig:MOT}\,(a) shows $N$ and $T$ as a function of the cooling light detuning $\delta_{cool}$. From these measurements we determine $\delta_{cool} = -8.6\, \Gamma$ ($\sim -50$\,MHz) and $\delta_{rep} = -8.4\, \Gamma$ as the optimal values.

The power of each MOT beam is about $P = 33$\,mW for each frequency, whose intensity $I_{MOT} = 4P/\pi D^2 \simeq 7.9$\,mW/cm$^2$ is well above the saturation intensity of this transition ($I_s^{D_2} = 2.54$\,mW/cm$^2$). The quadrupole magnetic field of the magneto-optical trap is generated by the coils in anti-Helmholtz configuration described in Section~\ref{sec:mag_MOT}. We also determine the optimal parameters of this field by maximizing the number of atoms in the sample while keeping its temperature as low as possible. Figure~\ref{fig:MOT}\,(b) shows a measurement of $N$ and $T$ as a function of the axial gradient of the quadrupole field, showing that the value $\partial B_z(z)/\partial z |_{z=0} \simeq 28$\,G/cm is optimal.

As a result, after a loading time of 8.6\,s we manage to capture up to $N = 5 \times 10^9$ atoms in the MOT at a temperature, still relatively high, of $T = 7$\,mK and atomic density of $n = 7.5 \times 10^{10}$ atoms/cm$^3$. The phase space density of the system is still very low, of the order of $\mbox{PSD} = n\lambda^3_{dB} = 4.7 \times 10^{-8}$, where $\lambda_{dB} = h/\sqrt{2\pi m k_B T}$ is the thermal de Broglie wavelength. In these measurements, as well as in all those presented in this paper, the temperature is obtained using the time-of-flight technique \cite{KetterleBose}.

		\subsubsection{Doppler and sub-Doppler cooling}\label{sec:subDopp}

In order to further cool down the sample and increase its phase space density, the gas undergoes two different additional laser cooling processes. We first apply an optical molasses cooling process based on the $D_2$ laser line that allows approaching the Doppler limit temperature \cite{Metcalf}. Next, we implement a gray-molasses technique, employing the $D_1$ line transitions to reach sub-Doppler temperatures \cite{Rio,Grier,Burchianti}. We provide details in the two following sections.

\paragraph*{\textbf{D$_2$ optical molasses cooling:}}\label{sec:D2Dopp}

The theoretical Doppler temperature limit for our sample is given by $T_D = \hbar \Gamma / 2k_B = 140.9 \,\mu$K. To reach this limit it is necessary to lower the intensity of the MOT light to minimize light-scattering heating, so the MOT light intensity should be much lower than the saturation intensity $I_s^{D_2}  = 2.54$\,mW/cm$^2$. Also, the cooling light must be detuned near to resonance, having an optimal value at $\delta_{cool} = -\Gamma/2$. The process needs to be done in absence of any magnetic field.

\begin{figure}
    \centering
    \includegraphics[width=\linewidth]{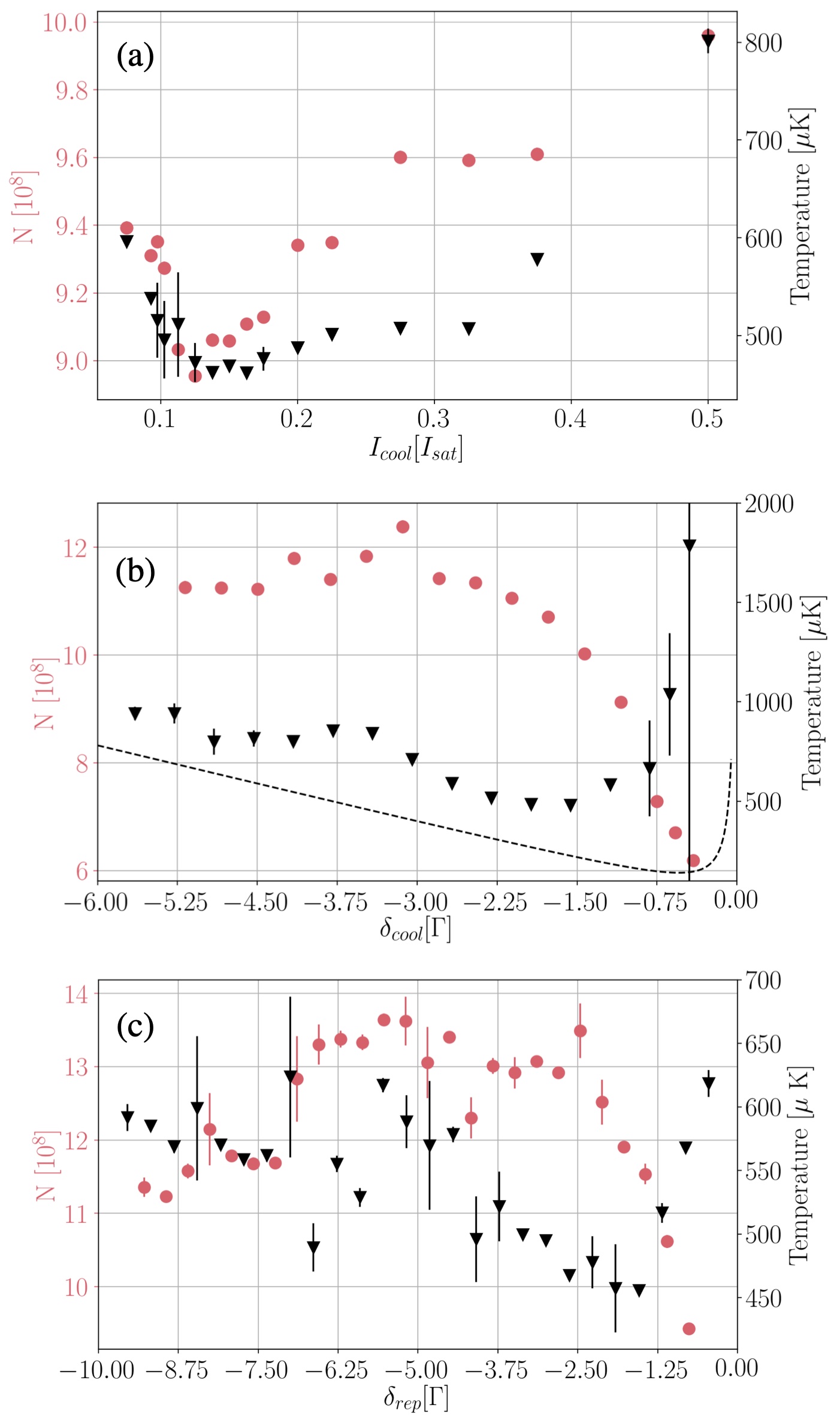}
    \caption{Number of atoms $N$ (red dots) and temperature $T$ (black triangles) of the atoms of the MOT after the D$_2$ optical molasses as a function of (a) the intensity of the cooling light and the detuning of (b) the cooling light and (c) the repumper light. The dashed black curve in (b) corresponds to the theoretical Doppler limit for the temperature of our sample. In these plots, the error bars correspond to one standard deviation of ten independent measurements. }
    \label{fig:D2}
\end{figure}

After loading the MOT we abruptly switch off the quadrupole magnetic field (we also switch off the Zeeman slower magnetic field 400\,ms before to guarantee the absence of any magnetic field in the sample region). Simultaneously, we decrease the intensity of the MOT beams and shift the value of cooling and repumper frequencies  towards resonance. Figure~\ref{fig:D2}\,(a) shows the effect on $N$ and $T$ of the intensity reduction, while Figures~\ref{fig:D2}\,(b) and (c) present the corresponding effect of the frequency shift of both MOT frequencies. 

As we can see, an important temperature drop is observed when the intensity of the light decreases. Concerning the frequency shift, as long as we keep the detuning below $-2\,\Gamma$, the number of atoms remains approximately constant while temperature decreases. We determine that the best values for intensity are $I_{cool} \simeq 0.35 I_s^{D_2}$ for cooling light and $I_{rep} \simeq 0.3 I_s^{D_2}$ for repumper, while the optimal frequency detuning is $\delta_{cool} = \delta_{rep} = -2\,\Gamma$. We also found that the optimal duration of this molasses process is 850\,$\mu$s; if shorter, the temperature does not reach the minimum possible value, and if longer we start losing atoms.

Under these conditions, we are able to cool down about $6 \times 10^8$ atoms to a temperature of about 500$\mu$K. The dashed black curve in Figure~\ref{fig:D2}\,(b) shows the theoretical Doppler limit, compared to which our experimental points lie above for the entire range of the detuning of cooling light considered. In other elements, such as as rubidium or cesium, it is observed not only that the Doppler limit is reached but even sub-Doppler temperatures are attained due to the emergence of the Sisyphus sub-Doppler cooling mechanism \cite{Dalibard}. For lithium this molasses scheme is not very efficient because the hyperfine levels of the state $2^2P_{3/2}$ cannot be well resolved, since their separation is smaller than $\Gamma$. This limits the efficiency of the cooling process and keeps the sample well above the Doppler limit. The increase of the phase space density is also not very good, and we improve only by a factor of 2, being of the order of $\mbox{PSD} = 1 \times 10^{-7}$. For this reason, we apply a second laser cooling technique that uses the $D_1$ line transitions, known as gray molasses, that allows true sub-Doppler cooling \cite{Grier,Burchianti}.

\paragraph*{\textbf{D$_1$ gray molasses sub-Doppler cooling:}}\label{sec:D1subDopp}	
	
Gray-molasses cooling is a two-photon process in $\Lambda$ configuration (see Figure~\ref{fig:D1D2}) which combines both, Sisyphus cooling \cite{Dalibard} and Velocity Selective Coherent Population Trapping (VSCPT) \cite{Aspect} as cooling mechanisms. More details can be seen in references \cite{Burchianti,Grynberg}. In few words, the cooling process occurs in the following way. On the one hand, the $\Lambda$ scheme creates two coherent states, a so-called ``bright state'' that interacts with the light fields and a ``dark state'' which doesn't. The transition probability from the dark to the bright state depends on the square of the momentum of the atoms, having as a consequence that the slowest atoms accumulate in the dark state. In other words, we have a velocity selective process that protects the slowest atoms from light-assisted heating. 

On the other hand, the $D_1$ light gets to the atoms through the same optical fibers used to produce the MOT (see Section~\ref{sec:optics} ), and hence they generate a 3D  polarization gradient. This allows a Sisyphus-like cooling scheme between bright and dark states which decreases the momentum of the atoms. In this way, while the Sisyphus cooling mechanism decreases the momentum of the atoms of the gas, the VSCPT process accumulates the slowest atoms in a dark state. This significantly decreases the temperature of the sample. 

In our experiment, we implement this cooling stage immediately after the $D_2$ molasses stage. We specifically use the $D_1$ transition frequencies $2^2S_{1/2} (F=3/2) \rightarrow 2^2P_{1/2} (F'=3/2)$, which we call ``cooling'' frequency, and $2^2S_{1/2} (F=1/2) \rightarrow 2^2P_{1/2} (F'=3/2)$, which we call ``repumper''. This nomenclature is inherited from the standard molasses. Both frequencies are blue detuned, the cooling frequency by $\delta_1$ and the repumper light by $\delta_2$. Another important parameter is the difference between these detunings that we define as $\delta = \delta_1 - \delta_2$. 

\begin{figure}
    \centering
    \includegraphics[width=\linewidth]{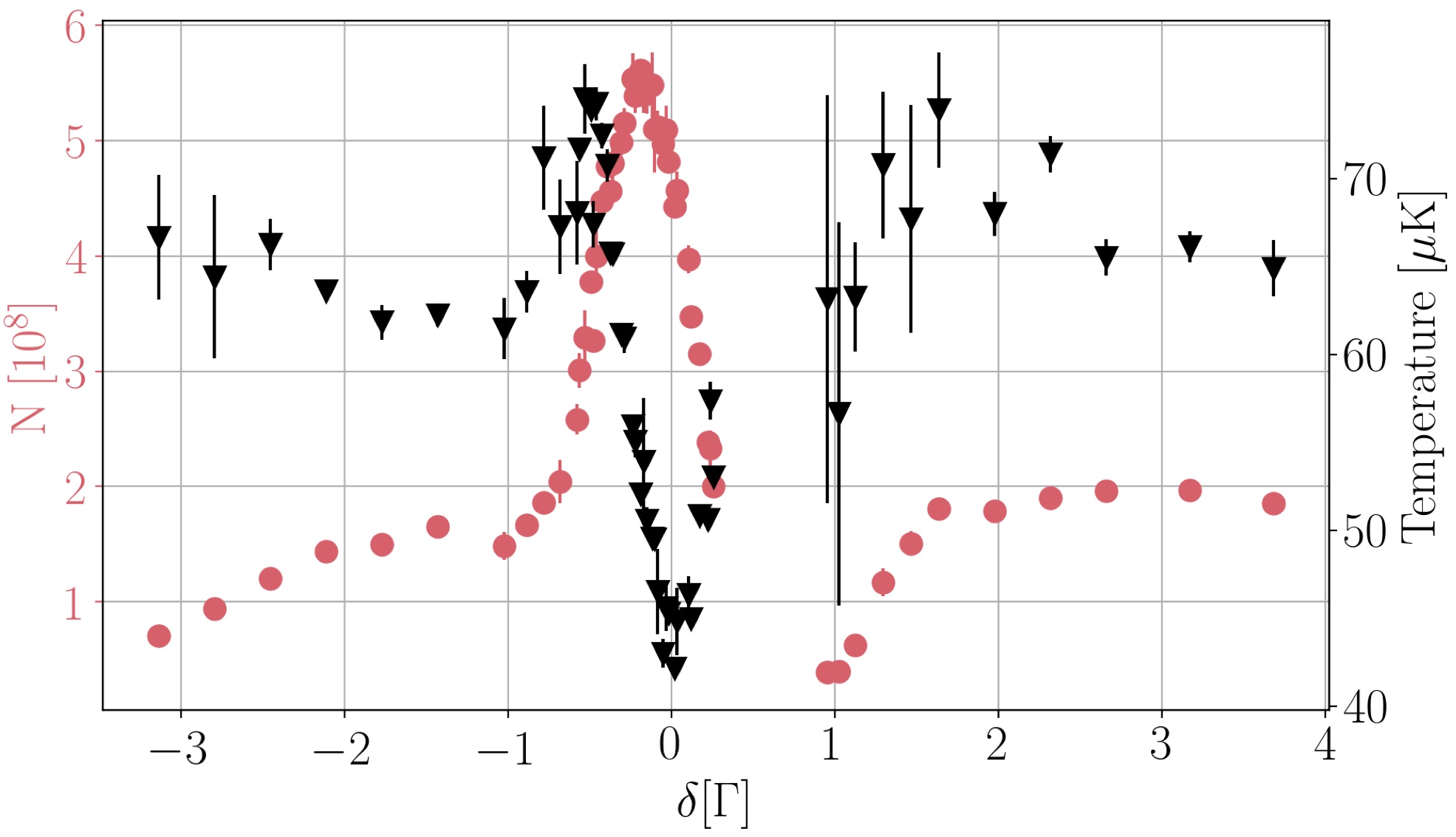}
    \caption{Number of atoms (red dots) and temperature (black triangles) of the sample as a function of the detuning between cooling and repumper light during gray molasses sub-Doppler cooling stage. The error bars correspond to one standard deviation of ten independent measurements}
    \label{fig:Fano}
\end{figure}

To characterize the gray-molasses we start by fixing $\delta_1 = +5.7\,\Gamma$ and keeping the repumper intensity low, at about $I_{rep} = 0.06I_s^{D_1}$, and the cooling intensity at its maximum value of the order of $I_{cool} \simeq I_s^{D_1}$. The saturation intensity for the $D_1$ line is $I_s^{D_1} = 7.59$\,mW/cm$^2$. We then measure the number of atoms and the temperature of the sample as the detuning difference $\delta$ varies. The results are shown in Figure~\ref{fig:Fano}.

We can see that the temperature follows a Fano-like profile, reaching a minimum at $\delta = 0$ (i.e. at $\delta_2 = \delta_1$), the so called Raman condition, in which the temperature is as low as $40\,\mu$K. Although the number of atoms does not reach its maximum at the Raman condition but at $\delta \approx -0.25\,\Gamma$, we still have a very good efficiency of the process at $\delta = 0$, being able to cool about 75\% of the atoms. These results are expected, as previously reported for the case of $^{6}$Li \cite{Burchianti}, and other atomic species such as $^{40}$K \cite{Rio} and $^{7}$Li \cite{Grier}. Notice that the plot of Figure~\ref{fig:Fano} has no data points in the interval $0.4 < \delta < 0.8$, as explained in reference \cite{Burchianti}, in this range the energy of the dark state becomes larger than the energy of the bright state and in consequence the VSCPT process significantly heats the cloud. In this range, the temperature becomes so high that time-of-flight measurements become very difficult to analyze and the measurement of $N$ and $T$ cannot be performed. Notice how the error bars of the data around that range consistently increase.

We also measure the effect of changing the cooling detuning $\delta_1$ while keeping the Raman condition $\delta = 0$. Both the number of atoms and the temperature remain constant in a wide interval of frequencies, showing the robustness of the gray-molasses process. We chose $\delta_1 = +5.7\,\Gamma$ for it is the value at which our acousto-optic modulators attain maximum efficiency.

The duration of the gray molasses is also an important parameter. We observe that after $400\,\mu$s, the efficiency of the process becomes nearly constant and better results are obtained for a duration time of  1\,ms. 

For the next stages, it is important to have all the atoms of the sample in the $F=1/2$ hyperfine state of the ground state $2^2S_{1/2}$ because the Feshbach resonance that we will use is present between its two magnetic sublevels. To do so, we switch off the $D_1$ repumper light $50\,\mu$s before the $D_1$ cooling light, so we manage to concentrate nearly 95\% of the atoms in the $F = 1/2$ hyperfine level.

To summarize, after the whole laser cooling process, we are able to produce a sample containing about $4.5 \times 10^8$ atoms in the hyperfine $F=1/2$ state at a temperature of $40~\mu$K. The phase space density increased considerably to $\mbox{PSD} \simeq 6.6 \times 10^{-6}$. This represents an excellent starting point for the subsequent cooling stages. 

Table~\ref{tab:parameters} presents the list of all the parameters employed in the laser cooling process.

\begin{table}
\caption{Optimized parameters of the optical cooling stages.\label{tab:parameters}}
\begin{tabular}{c c c}
\hline \\ [-2mm]
\textbf{Cooling stage} 						& \textbf{Parameter} 							& \textbf{Optimal Value} 		\\ [1mm] \hline  \\ [-2mm]
\multirow{7}{*}{MOT}							& $\partial B_z(z)/\partial z |_{z=0}$	& 28\,G/cm							\\ [1mm] 
															& $\delta_{cool}$									& -8.6 $\Gamma$				\\ [1mm]
															& $\delta_{rep}$									& -8.4 $\Gamma$				\\ [1mm]
															& Loading time										& 8.6\,s								\\ [1mm]
															& N														& $5 \times 10^9$ atoms	\\ [1mm]
															& T														& 7\,mK								\\ [1mm]
															& PSD													& $4.7 \times 10^{-8}$		\\ [1mm] \hline \\[-2mm]
\multirow{8}{*}{D$_2$ Molasses}			& $\delta_{cool}$                  				& -2 $\Gamma$					\\ [1mm]
															& $\delta_{rep}$                  				& -2 $\Gamma$					\\ [1mm]
															& $I_{cool}$                  						& 0.35 $I_s^{D_2}$				\\ [1mm]
															& $I_{rep}$                  							& 0.30 $I_s^{D_2}$				\\ [1mm]
															& Duration                  							& 850\,$\mu$s					\\ [1mm]
															& N														& $6 \times 10^8$ atoms	\\ [1mm]
															& T														& 500\,$\mu$K					\\ [1mm]
															& PSD													& $1 \times 10^{-7}$			\\ [1mm] \hline \\[-2mm]
\multirow{8}{*}{D$_1$ Gray molasses}	& $\delta_{1}$										& +5.7 $\Gamma$				\\ [1mm]
															& $\delta_{2}$										& +5.7 $\Gamma$				\\ [1mm]
															& $I_{cool}$											& $I_s^{D_1}$						\\ [1mm]
															& $I_{rep}$											& 0.06 $I_s^{D_1}$				\\ [1mm]
															& Duration											& 1\,ms								\\ [1mm]
															& N														& $4.5 \times 10^8$ atoms	\\ [1mm]
															& T														& 40\,$\mu$K						\\ [1mm]
															& PSD													& $6.6 \times 10^{-6}$		\\ [1mm] \hline
\end{tabular}
\end{table}

	\subsection{Cooling the sample to quantum degeneracy}\label{sec:evaporation}

After the $D_2$ and $D_1$ cooling stages, the sample is ready to be transferred into a conservative potential in which evaporative cooling can be applied and quantum degeneracy is achieved. In the following sections we explain how this process is done in our setup.

		\subsubsection{Transference into the conservative trap}\label{sec:transfer_ODT}

As explained in Section~\ref{sec:ODT}\, , our trap is created as the composition of a single-beam optical dipole trap and a magnetic curvature, which provide, respectively, radial and axial confinement.

During the $D_1$ cooling process we ramp the power of the optical dipole trap (ODT) to 160\,W in 7\,ms. The beam is focused right at the center of the atomic cloud, as shown in Figure~\ref{fig:Beams}. Once the power of the optical beam has reached its maximum value, we ramp the Feshbach magnetic field to 832\,G in 50\,ms. This field corresponds to the unitary limit in which the scattering length diverges, which is optimal for the following evaporative cooling stage because the collision rate is maximized and the thermalization process is optimized. 

When the magnetic field is ramped up, the $F=1/2$ hyperfine state splits into the two states $| 1 \rangle$ and  $| 2 \rangle$, where $| 1 \rangle$ has lowest energy for all magnetic fields. In the magnetic fields that we employ these states are well within the Paschen-Back regime, so the energy difference between them remains almost unchanged. Moreover, if the ramp of the magnetic field is adiabatic, both states are nearly equally populated, so we create a well balanced mixture.

The Feshbach field curvature provides an axial harmonic confinement of about $\omega_{z_{mag}} = 2\pi \times 11$\,Hz. This confinement, of course, is negligible at the beginning of the ODT loading since at such high power the confinement provided by the optical trap is much higher, $\omega_{r_{ODT}} \simeq 2\pi \times 10$\,kHz and $\omega_{z_{ODT}} \simeq 2\pi \times 87$\,Hz (see Eq.~(\ref{eq:hybrid_freqs})), however, the magnetic confinement becomes more and more important as we apply the evaporative cooling process in which the power of the ODT laser beam is gradually decreased.

\begin{figure}
    \centering
    \includegraphics[width=\linewidth]{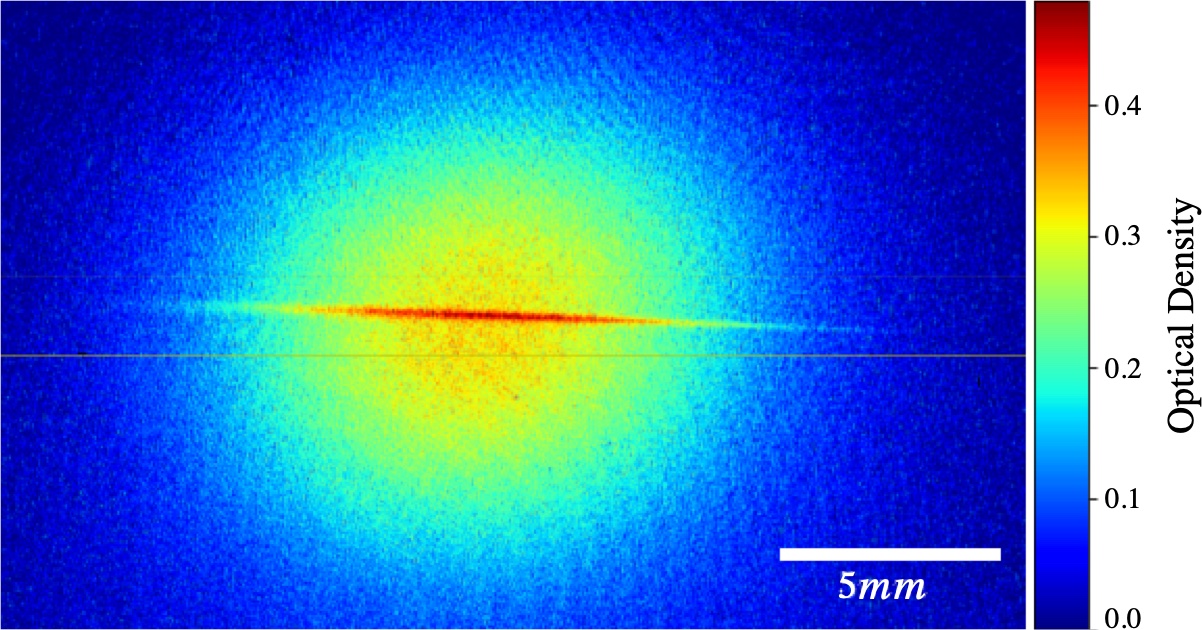}
    \caption{Absorption imaging of the atoms transferred to the optical dipole trap (horizontal darker region) from the laser sub-Doppler cooled sample (round lighter region). The color gradient corresponds to the optical density of the sample according to the color bar on the right.}
    \label{fig:ODT}
\end{figure}

After the optical and magnetic fields have been ramped up, we trap about $3 \times 10^6$ atoms in the conservative potential, which means that our trapping efficiency is of the order of 1\%. We hold the atoms in this trap for 20\,ms to let them settle in the minimum of the potential. At this point we can implement the evaporative cooling process \cite{Ketterle}, which is the last step before reaching quantum degeneracy.  Since the trap increases the density of the sample, we observe a considerable increase of the temperature of the sample to about $200\,\mu$K. Figure~\ref{fig:ODT} shows an absorption image of the atoms from the sub-Doppler cooled sample transferred into the ODT beam.

		\subsubsection{Evaporative cooling}\label{sec:evap_cool}
	
Evaporative cooling is performed by ramping down the ODT power while keeping the magnetic field at 832\,G. To achieve runaway evaporation it is fundamental that the collision rate does not decrease as the atoms are evaporated, this means that the density of the cloud needs to increase as its temperature is reduced. To guarantee this condition, the evaporation process must be performed slow enough for thermalization to occur. At the same time, the evaporation has to be the main loss process, so it cannot be too slow for the background-vapor collisions with the sample to be important. A good quantity to evaluate the effectiveness of the evaporation process is the phase space density, $\mbox{PSD} = n\lambda^3_{dB} \propto n/T^{3/2}$, which must increase as the evaporation is applied \cite{Ketterle}.

\begin{figure}
    \centering
    \includegraphics[width=\linewidth]{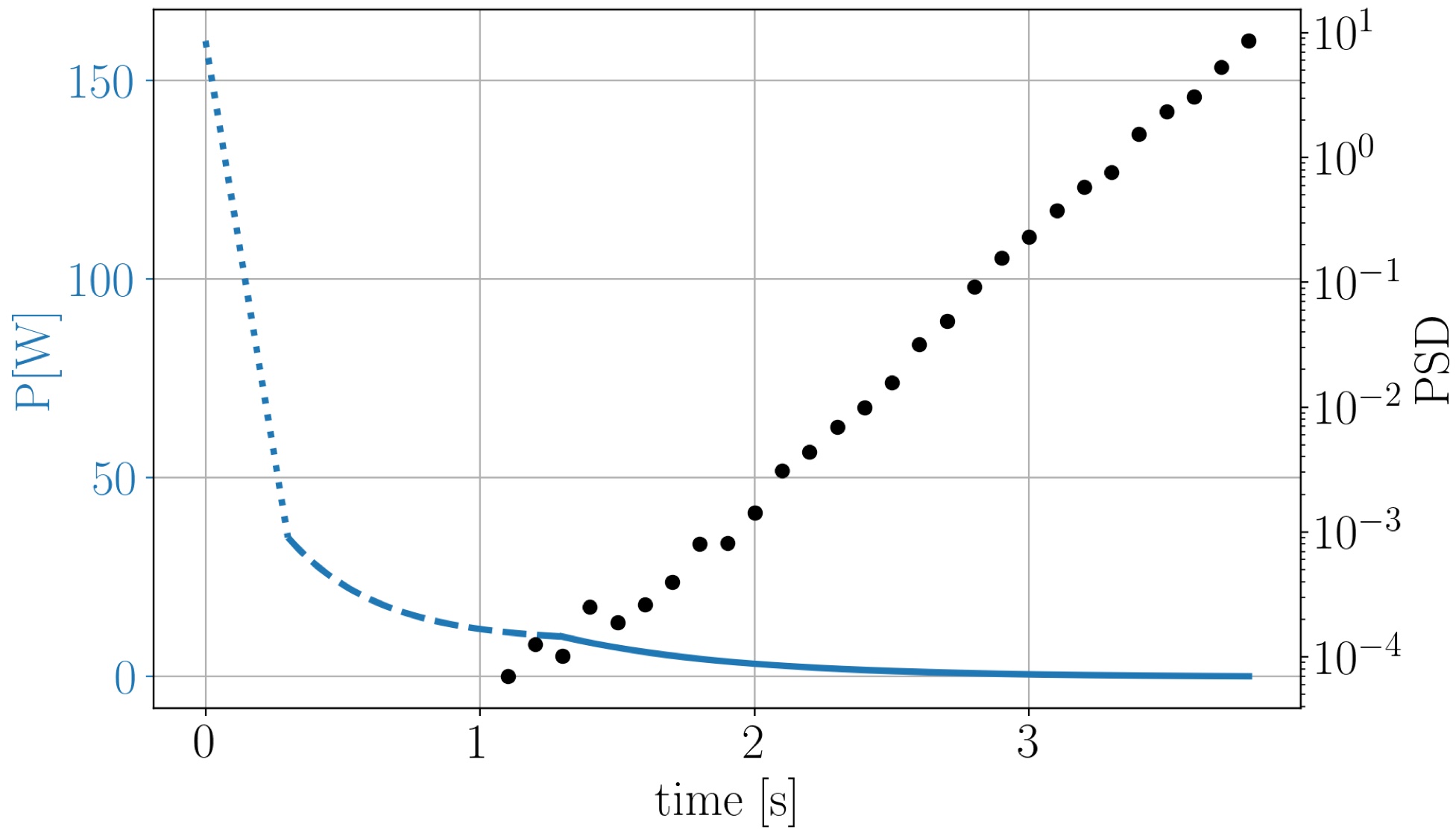}
    \caption{Blue curves: Plot of the evaporation ramps performed by decreasing the power of the optical dipole trap (not a measurement), see text for details. Black data points: Measurement of the phase space density of the system during evaporation. For these measurements, the uncertainty is of the order of 10\%, corresponding to one standard deviation of ten independent measurements, however, the error bars are not visible at the scale of the graph.}
    \label{fig:Evap}
\end{figure}

The evaporative process is performed by concatenating three exponential ramps, as shown in the blue curves of Figure~\ref{fig:Evap}. The first ramp goes from 160\,W to 35\,W in 300\,ms having a characteristic time of $\tau_1 = 125$\,ms (dotted curve in Fig.~\ref{fig:Evap}); the second ramp, from 35\,W to 10\,W in 1.0\,s, with $\tau_2 = 440$\,ms  (dashed curve), and finally, a very slow ramp from 10\,W to a variable value of the order of $P_0 = 35$\,mW in 2.6\,s, with $\tau_3 = 2000$\,ms  (solid curve). The total duration of the evaporation process is 3.8\,s. These parameters are determined by maximizing the phase density of the system. The black data points in Figure~\ref{fig:Evap} shows how the measured $\mbox{PSD}$ increases as the evaporation proceeds. Notice that $\mbox{PSD} \geq 1$ at the end of the last ramp, indicating the onset of quantum degeneracy.

At the end of the third evaporation ramp we adiabatically ramp the Feshbach field to the corresponding value in order to produce a sample in any desired interaction regime across the Feshbach resonance; this magnetic ramp lasts about 300\,ms. The regimes that we are interested in exploring are within the interval of 670 to 900\,G, which contains the BEC-BCS crossover. 

By changing the value of the Feshbach field, we also modify the curvature of the magnetic field; however, it changes less than a 10\% within the mentioned interval of interest, which means that we do not significantly modify the geometry of the trap as we change the scattering length. Of course, as can be seen from Equation~(\ref{eq:hybrid_freqs}), the frequencies of the trap depend on the power of the ODT, which in turn, determines the temperature and degree of degeneracy of the sample.

After the evaporative cooling process we are able to produce quantum degenerate superfluid samples containing about $N = 5 \times 10^4$ atomic pairs at a temperature of the order of $T/T_F = 0.1$ (which corresponds for this value of $N$ to approximately 20\,nK) and a phase space density well above the unity, of the order of $\mbox{PSD} \approx 10$, demonstrating the fully degenerate nature of our sample. The trap frequencies are $\omega_r = 2\pi \times 163$~Hz and $\omega_z = 2\pi \times 11$ Hz, which means that our sample is cigar-shaped with an aspect ratio of the order of $1:15$. The duty cycle of our experiment is shorter than 14s.

		\subsubsection{Superfluids across the BEC-BCS crossover}\label{sec:BEC_BCS}

As mentioned in the previous section, we select the interacting regime of the produced sample at the end of the last evaporation ramp by means of the Feshbach resonance that allows us to set the value of the scattering length $a_s$. As explained in Section~\ref{sec:diag_optics}\,, we are able to produce and probe samples at practically any magnetic field up to 1200\,G. Specifically, as we explain below, we are able to produce ultracold  superfluid samples within the interaction range of $-0.65 \leq (k_F\,a_s)^{-1} \leq 7.6$, which means that we can produce samples from the deep (weakly interacting) BEC regime to the strongly interacting BCS regime, passing, of course, through unitarity at $(k_F\,a_s)^{-1} = 0$. Clearly, we have access to most of the crossover region, $-1 \leq (k_F\,a_s)^{-1} \leq 1$, corresponding to the magnetic field interval 790\,G to 900\,G. 

\begin{figure}
    \centering
    \includegraphics[width=\linewidth]{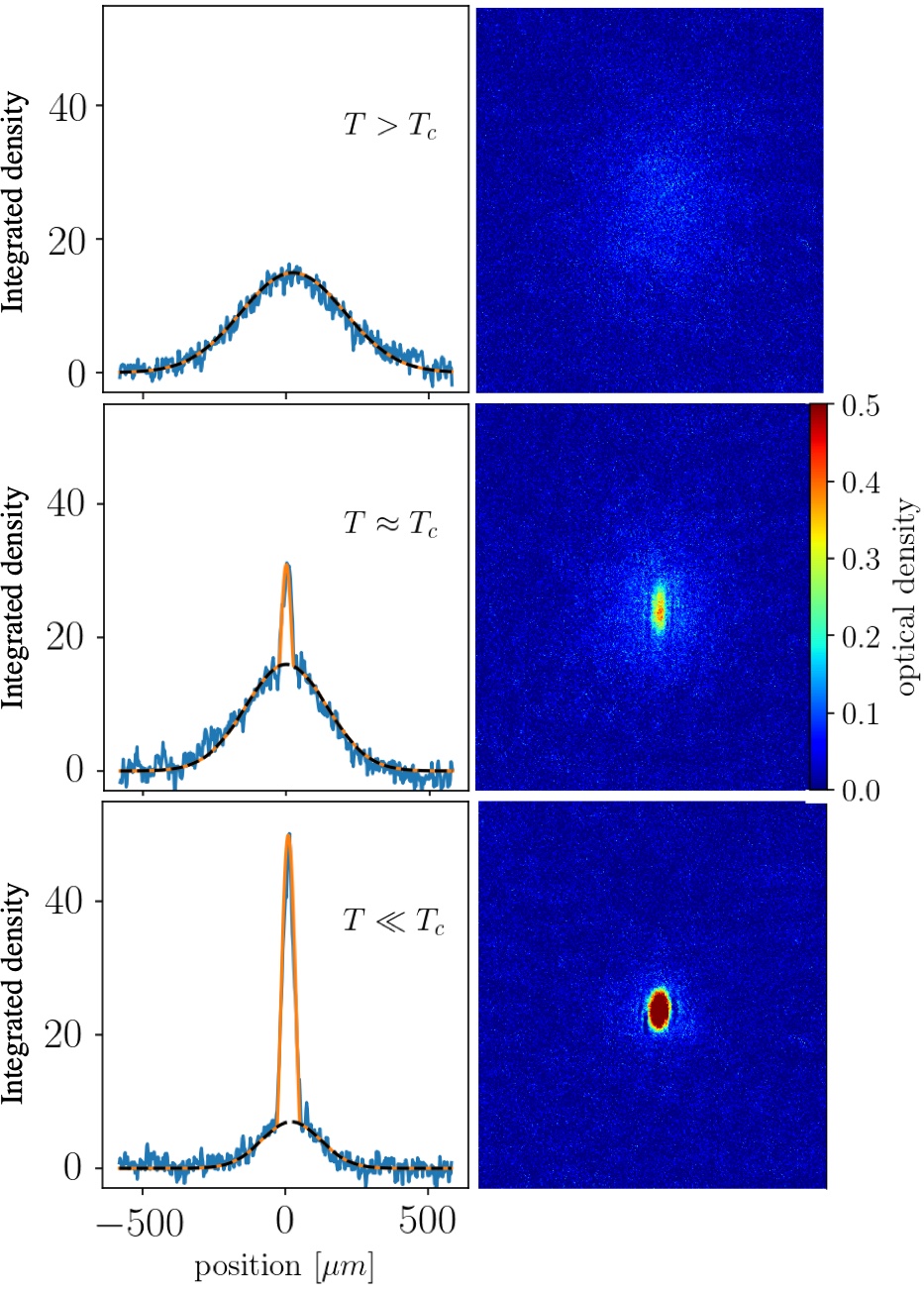}
    \caption{Absorption images of the atomic samples (right pictures) and their corresponding integrated density profile (left graphs) as temperature is decreased. Upper panels: thermal gas above critical temperature $T_C$. Middle panels: gas just below the critical temperature, notice the bimodal gaussian-parabolic distribution. Lower panels: molecular Bose-Einstein condensate well below the critical temperature, the parabolic distribution is dominant and the gaussian one is barely noticeable. The color gradient corresponds to the optical density of the gas. All pictures were taken after a time-of-flight of 15\,ms. In the graphs, the dashed black line corresponds to a fitting of only the gaussian wings, while the orange solid line to the bimodal distribution.}
    \label{fig:BEC}
\end{figure}

\begin{figure*}
    \centering
    \includegraphics[width=0.9\linewidth]{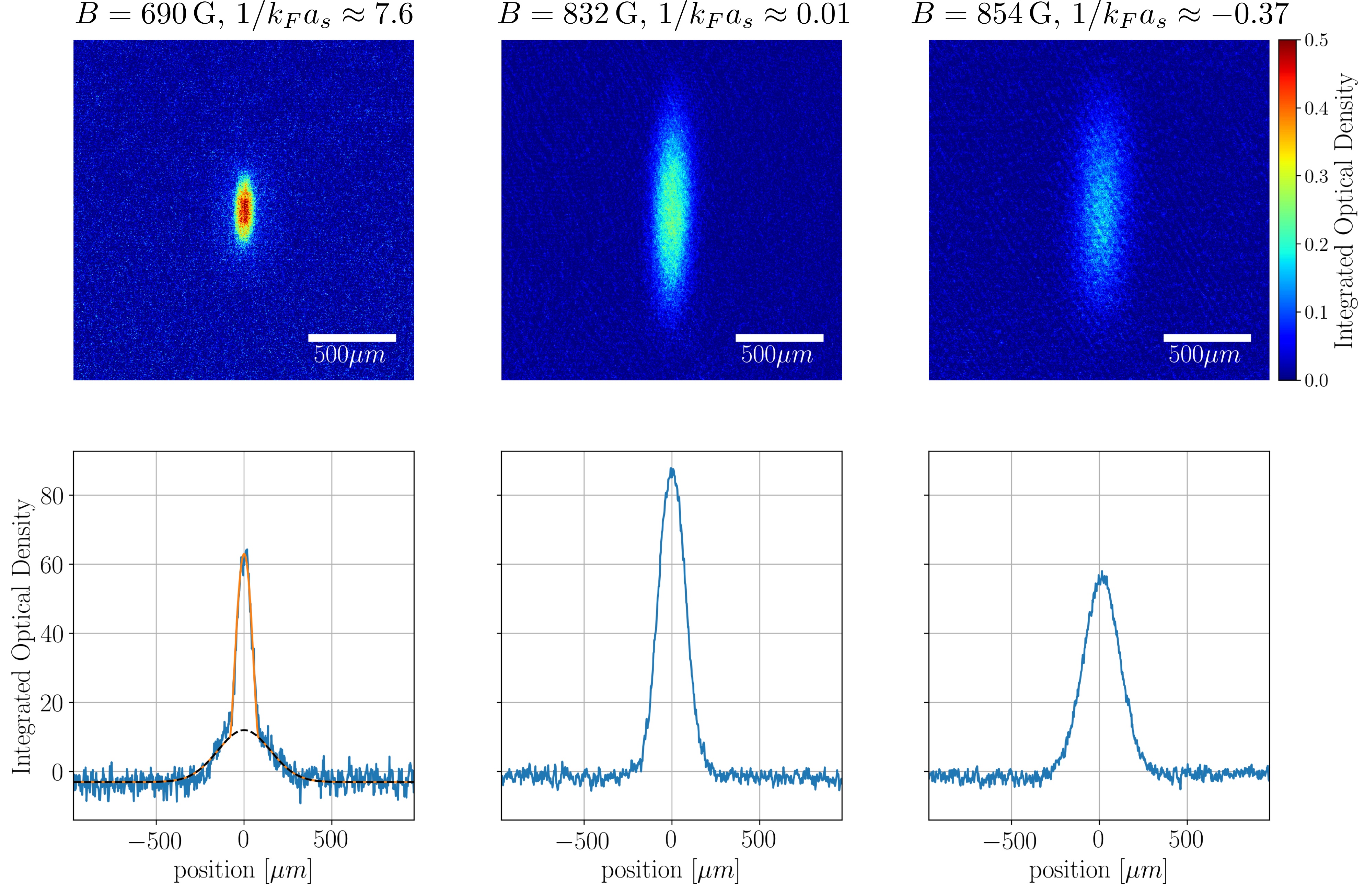}
    \caption{Absorption images of quantum degenerate atomic samples (upper pictures) and their corresponding integrated density profile (lower graphs) as the scattering length is varied across the BEC-BCS crossover. Left panels: Bose-Einstein condensate of molecules at $(k_F\,a_s)^{-1} \approx 7.6$, the bimodal and gaussian fits are shown as a solid orange and black dashed lines, respectively. Middle panels: superfluid gas at unitarity at $(k_F\,a_s)^{-1} \approx 0.01$. Right panels: ultracold gas at the BCS side of the Feshbach resonance at at $(k_F\,a_s)^{-1} \approx -0.37$. The color gradient corresponds to the optical density of the gas. All pictures were taken after a time-of-flight of 20\,ms. }
    \label{fig:BEC-BCS}
\end{figure*}

Evidently, the most important point here is to achieve, at every interacting regime, temperatures that are below the critical superfluid temperature, $T_C$. On the deep BEC side, $(k_F\,a_s)^{-1} > 1$, the critical temperature is approximately $T^{\mbox{\tiny{ BEC}}}_{C} \simeq 0.52 T_F$ and it is nearly independent of the scattering length \cite{Pethick, Giorgini96}. The minimum temperature attainable in our experiment, $T/T_F = 0.1$, remains  well below $T^{\mbox{\tiny{ BEC}}}_{C}$. In this case, the density profile of the cloud exhibits the very characteristic bimodal distribution \cite{Pethick}. The condensed fraction presents a parabolic sharp density profile that arises from the Thomas-Fermi approximation, while the non-condensed thermal atoms follow a gaussian Maxwell-Boltzmann distribution, which we use to estimate the temperature of the cloud in time-of-flight (TOF) imaging \cite{KetterleBose}. These features can be seen in Figure~\ref{fig:BEC}. The weakest interacting BEC that we can produce corresponds to a magnetic field of 670\,G for which $a_s = 1080\,a_0$ and $(k_F a_s)^{-1} = 7.6$. For lower magnetic fields the lifetime of the molecular condensate is too short to perform any typical experiment (it is shorter than 100\,ms, while in any other regime described here, it is of the order of 1.5\,s).

As the scattering length increases, within the BEC-BCS crossover range, and specially right at unitarity, this well defined bimodal distribution starts to wash out and becomes broader due to strong interactions \cite{Giorgini,Regal,Perali,He}. In this regime, it is not possible to discriminate between the superfluid fraction and the thermal fraction, and the density profile looks nearly Gaussian. However, we know that we are in the superfluid regime due to the following consideration. On the vicinity of the unitary limit the critical temperature is given by $T^{\mbox{\tiny{U}}}_{C} \simeq 0.167 T_F$ \cite{Ku}, which again, is above the temperature of our sample.

In contrast, on the BCS side of the crossover, the critical temperature is given by \cite{Giorgini, Sa}:
\begin{eqnarray}\label{eq:TCBCS}
T^{\mbox{\tiny{ BCS}}}_{C} \simeq 0.28 T_F \, e^{-\pi/2k_F \left| a_s \right|},
\end{eqnarray}
so it exponentially decays as the quantity $\left| k_F\,a_s \right|^{-1}$ increases. For instance, at $\left( k_F\,a_s \right)^{-1} = -0.65$, the critical temperature for the superfluid state is $T^{\mbox{\tiny{ BCS}}}_{C} / T_F \approx 0.1$, which is comparable to the minimum achievable temperature of our setup. In consequence, we cannot access the deep (weakly interacting) BCS superfluid regime because the critical temperature is below the technical limit of our experiment. This means that in our setup, superfluid regimes are attainable within the range $-0.65 \leq (k_F\,a_s)^{-1} \leq 7.6$. Figure~\ref{fig:BEC-BCS} shows a sequence of absorption images of a superfluid at $T/T_F = 0.1$ containing $N = 5  \times 10^4$ atomic pairs, as the scattering length changes from the BEC to the BCS regimes across the crossover. 

Besides the considerations concerning the critical temperature that we have presented here, we have also performed an additional measurement that ensures that all the observed regimes present superfluidity. Right after releasing the atoms from the trap, we have performed a fast Feshbach magnetic field ramp from the strongly interacting regimes into the deep BEC side \cite{Regal04, Zweirlein04}. As result of this ramp, the many-body wave function of the system is projected onto the far BEC side of the resonance. In all cases we observe the characteristic BEC bimodal distribution in the density profile, indicating that at unitarity and its vicinity we always have condensation of atomic pairs.

\section{Conclusion and Future Perspectives}\label{sec:conclusions}

We have presented the experimental setup and methods we use to produce and study ultracold fermionic superfluid samples of $^6$Li. We are able to generate samples containing $5 \times 10^4$ atomic pairs at temperatures as low as $T/T_F = 0.1$ at any superfluid regime across the BEC-BCS crossover within a duty cycle shorter than 14\,s. Our setup combines versatility and state-of-art techniques, which will allow us to study different aspects of quantum matter.

As a future perspective, we plan to study topological and hydrodynamic excitations such as quantized vortices (see for instance \cite{Tsubota,Tsatsos} and references therein). We specifically want to understand how the dynamics of these systems depend on the interacting regime as well as on the temperature of the cloud. To carry out these experiments, we need to expand the capabilities of our imaging system. In particular, as a complementary technique to our current absorption imaging system, we will implement the non-destructive phase contrast imaging technique \cite{KetterleBose} that will allow us to perform several images of the same sample without perturbing it. This is very important to address the dynamics of the superfluid sample.

As a long term perspective, we plan to produce ultracold samples of $^7$Li, a bosonic stable isotope of lithium. This is possible because we have also placed purified $^7$Li in our oven. The optical frequencies of the $D_1$ and $D_2$ lines of $^7$Li are very close to those of fermionic $^6$Li \cite{Das}. This means that with minor modifications on the optical cooling setup we should be able to produce, alternatively, bosonic ultracold samples of $^7$Li. This is very interesting because this species also presents a broad Feshbach resonance, opening the possibility to study very weakly interacting bosonic systems, a regime that our current setup does not offer and which represents an excellent scenario to study the thermodynamic properties of the superfluid to normal gas transition.

\acknowledgments

Our experimental setup is the first apparatus able to produce quantum gases in Mexico, it is also the first one capable of producing fermionic ultracold systems in Latin America. We hope that our work helps to incentive the growth of this exciting and interesting research field in our country and in the Latin-American region. 

The construction of our experiment has represented a major challenge. This has only been possible thanks to the support of numerous institutions, colleagues and different funding sources. We would like to use this space to make the corresponding acknowledgments.

We want to thank our colleagues from Instituto de Física, UNAM:

\begin{itemize}
\item[--] To Manuel Torres Labansat, Víctor Romero-Rochín, Daniel Sahagún Sánchez, Rosario Paredes Gutiérrez, Asaf Paris Mandoki and Carlos Villarreal Luján for their unconditional support and fruitful discussions. 
\item[--] To Roberto Gleason Villagrán and Marco Antonio Veytia Vidana for their excellent work during the installation of the basic infrastructure of the laboratory.
\item[--] To the former members of our laboratory, Eduardo Ibarra García-Padilla, Jesús Ernesto Carro Martínez, Cristian Adán Mojica Casique, Iliana Isabel Cortés Pérez and Aurora Guadalupe Borges Sánchez,  for their contribution during the early stages of construction of the experimental setup. 
\item[--] To Giovanni Alonso Torres for his help in preparing Figure~\ref{fig:UHV}.
\end{itemize}

We also thank our colleagues from INO-CNR and LENS:
\begin{itemize}
\item[--] Alessia Burchianti and Chiara Fort for their technical advice in setting our automation and vacuum systems.
\item[--] Andreas Trenkwalder for his help in setting different analysis algorithms as well as his advice concerning our imaging setup.
\item[--] Marco de Pas for his advice in the construction of several electronic devices and instruments.
\item[--] In general, the entire Quantum Gases Group for their very useful advice.
\end{itemize}

We acknowledge as well Greg S. Engel (University of Chicago) for his advice in the installation of our air-conditioning system.

We are deeply grateful for the financial support provided by the following institutions:
\begin{itemize}
\item[--] CIC-UNAM and CONACyT through the National Laboratories program, with grants number: LN232652, LN260704, LN271322, LN280181, LN293471 and LN299057.
\item[--] CONACyT through Ciencia Básica grants 255573 and 254942.
\item[--] Insituto de Física UNAM through grants PIIF-8 and PIIF-9.
\item[--] DGAPA-UNAM-PAPIIT grant numbers IA101716, IN111516, IN107014 and IN103818.
\item[--] CONACyT through Redes Tématicas ``Tecnología Cuántica''.
\end{itemize}

F.J.P.C. would like to thank the support from DGAPA-UNAM and from CLAF-SeCyTI for former postdoctoral fellowships.  D.H.R., J.E.P.C., M.M.L., R.C.R., A.G.V. and S.A.M.R. acknowledge their scholarships from CONACyT, while D.H.R. also acknowledge a scholarship from DGAPA-UNAM-PAPIIT.

We finally want to thank the company Seman Baker S.A. de C.V. for their generous donation of numerous machined pieces.


\begin{thebibliography}{}
\bibitem{Fermi} The Fermi wave vector is defined by $k_F \equiv \sqrt{2mE_F}/\hbar$, where $E_F$ is the Fermi energy, which corresponds to the energy of a fermion at the Fermi surface. For the case of $N$ identical fermions trapped in a harmonic potential, the Fermi energy is given by $E_F = \hbar\bar{\omega}(6N)^{1/3}$, where $\bar{\omega} = (\omega_x\omega_y\omega_z)^{1/3}$ is the geometric average of the trap frequencies. From it, we can define other important quantities such as the Fermi temperature, $T_F \equiv E_F/k_B$, which establishes a temperature scale for the occurrence of quantum degeneracy.
\bibitem{flux} This flux can be roughly estimated with the formula $\phi = n\bar{v}A/4$ where $n= 1.31\times 10^{12}$ atoms/cm$^3$ is the atomic density in the oven, $\bar{v} = \sqrt{8k_BT/\pi m} \simeq 1,540$\,m/s is the mean velocity of the atoms, and $A = 12.56$\,mm$^2$ is the nozzle area.
\bibitem{a} This deceleration depends on the parameters of the cooling frequency of the Zeeman slower light as $a = \frac{\hbar k}{m} \frac{\Gamma}{2}\frac{s}{1+s}$, where $s = I/I_s$ is the saturation parameter, being $I$ the intensity of the light and $I_s = 2.54$\,mW/cm$^2$ the saturation intensity of the $D_2$ line. For our experimental conditions $s \simeq 10$, so we are close to the maximum attainable deceleration.
\end{thebibliography}
\end{document}